\documentclass[iop,apjl,12pt]{emulateapj}
\usepackage{times}
\usepackage[normalem]{ulem}
\usepackage{apjfonts}
\usepackage{epsfig}
\usepackage{natbib}
\usepackage{amsfonts}
\usepackage{amsmath}
\usepackage{multirow}
\usepackage{enumerate}
\usepackage[usenames]{color}
\usepackage[plainpages=false, colorlinks=true, anchorcolor=lblui, linkcolor=blue, citecolor=black, urlcolor=blue, bookmarks=false]{hyperref}
\usepackage{threeparttable}
\def\Dwa{$\,$\uppercase\expandafter{\romannumeral5}$\,$}

\def\sless{\lower2pt\hbox{$\buildrel {\scriptstyle <}
   \over {\scriptstyle\sim}$}}

\def\sgreat{\lower2pt\hbox{$\buildrel {\scriptstyle >}
   \over {\scriptstyle\sim}$}}
\def\sharpnull#1{}

\newcommand{\shortauth}{M{\"o}sta \emph{et al.}}
\newcommand{\slugcom}{Draft version - \today}
\slugcomment{\slugcom}

\lefthead{\sc \footnotesize \slugcom \hfill \shortauth}
\righthead{\sc \footnotesize \slugcom \hfill \shortauth}

\begin{document}
\slugcomment{Draft version \today.}

\title{R-process Nucleosynthesis from Three-Dimensional Magnetorotational Core-Collapse Supernovae}

\author{Philipp M{\"o}sta\altaffilmark{1+}}
\author{Luke F. Roberts\altaffilmark{2}}
\author{Goni Halevi\altaffilmark{1,3}}
\author{Christian D. Ott\altaffilmark{4,5}}
\author{Jonas Lippuner\altaffilmark{6,7,8}}
\author{Roland Haas\altaffilmark{9}}
\author{Erik Schnetter\altaffilmark{10,11,12}}
  \altaffiltext{1}{Department of Astronomy, 
                   501 Campbell Hall \#3411, \\University of California at Berkeley, 
                   Berkeley, CA 94720\\
                   pmoesta@berkeley.edu}
\altaffiltext{2}{National Superconducting Cyclotron Laboratory and Department of Physics, Michigan State University, East Lansing,USA}
\altaffiltext{3}{Department of Astrophysical Sciences, Princeton University, Princeton, NJ 08544, USA}
\altaffiltext{4}{TAPIR, California Institute of Technology, Pasadena, USA}
\altaffiltext{5}{Yukawa Institute for Theoretical Physics, Kyoto University, Kyoto, Japan}
\altaffiltext{6}{CCS-2, Los Alamos National Laboratory, P.O. Box 1663, Los Alamos, NM
87545, USA.}
\altaffiltext{7}{Center for Nonlinear Studies, Los Alamos National Laboratory, P.O.
Box 1663, Los Alamos, NM 87545, USA.}
\altaffiltext{8}{Center for Theoretical Astrophysics, Los Alamos National Laboratory,
P.O. Box 1663, Los Alamos, NM, 87545, USA.}
\altaffiltext{9}{NCSA, University of Illinois, Urbana-Champaign, USA.}
\altaffiltext{10}{Perimeter Institute for Theoretical Physics, Waterloo, ON, Canada.}
\altaffiltext{11}{Department of Physics, University of Guelph, Guelph, ON, Canada.}
\altaffiltext{12}{Center for Computation \& Technology, Louisiana State University, Baton Rouge, USA.}
\altaffiltext{+}{NASA Einstein Fellow}

\begin{abstract}
We investigate $r$-process nucleosynthesis in three-dimensional (3D)
general-relativistic magnetohydrodynamic simulations of rapidly rotating
strongly magnetized core collapse. The simulations include a microphysical
finite-temperature equation of state and a leakage scheme that captures the
overall energetics and lepton number exchange due to postbounce neutrino
emission and absorption. We track the composition of the ejected material using
the nuclear reaction network \texttt{SkyNet}. Our results show that the 3D
dynamics of magnetorotational core-collapse supernovae (CCSN) are important for
their nucleosynthetic signature. We find that production of $r$-process
material beyond the second peak is reduced by a factor of 100 when the
magnetorotational jets produced by the rapidly rotating core undergo a kink
instability. Our results indicate that 3D magnetorotationally powered CCSNe are
a robust $r$-process source only if they are obtained by the collapse of cores
with unrealistically large precollapse magnetic fields of order
$10^{13}\,\mathrm{G}$. Additionally, a comparison simulation that we restrict
to axisymmetry, results in overly optimistic $r$-process production for lower
magnetic field strengths. 

\end{abstract}

\section{Introduction}

\begin{figure*}[t]
\includegraphics[width=0.33\textwidth]{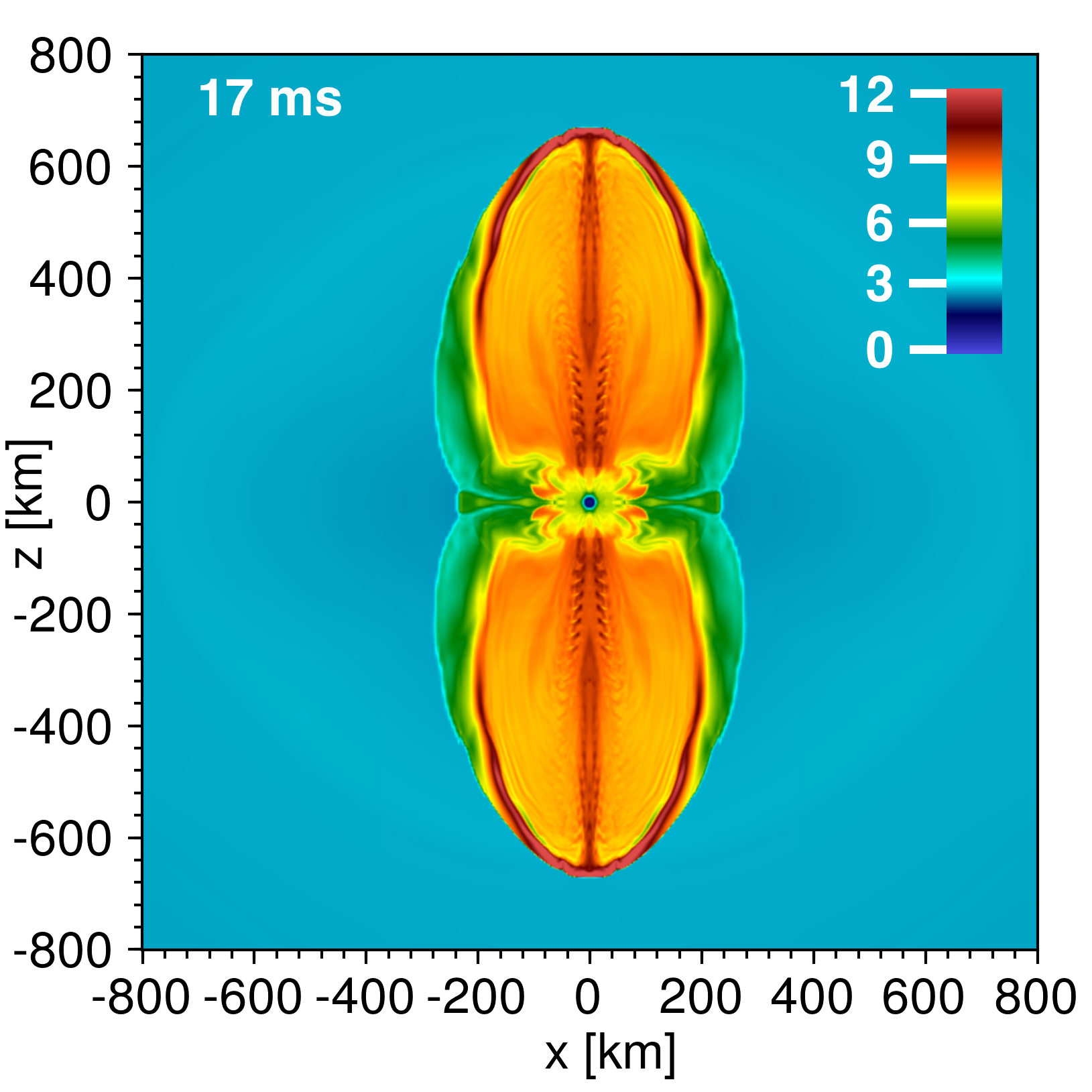}
\includegraphics[width=0.33\textwidth]{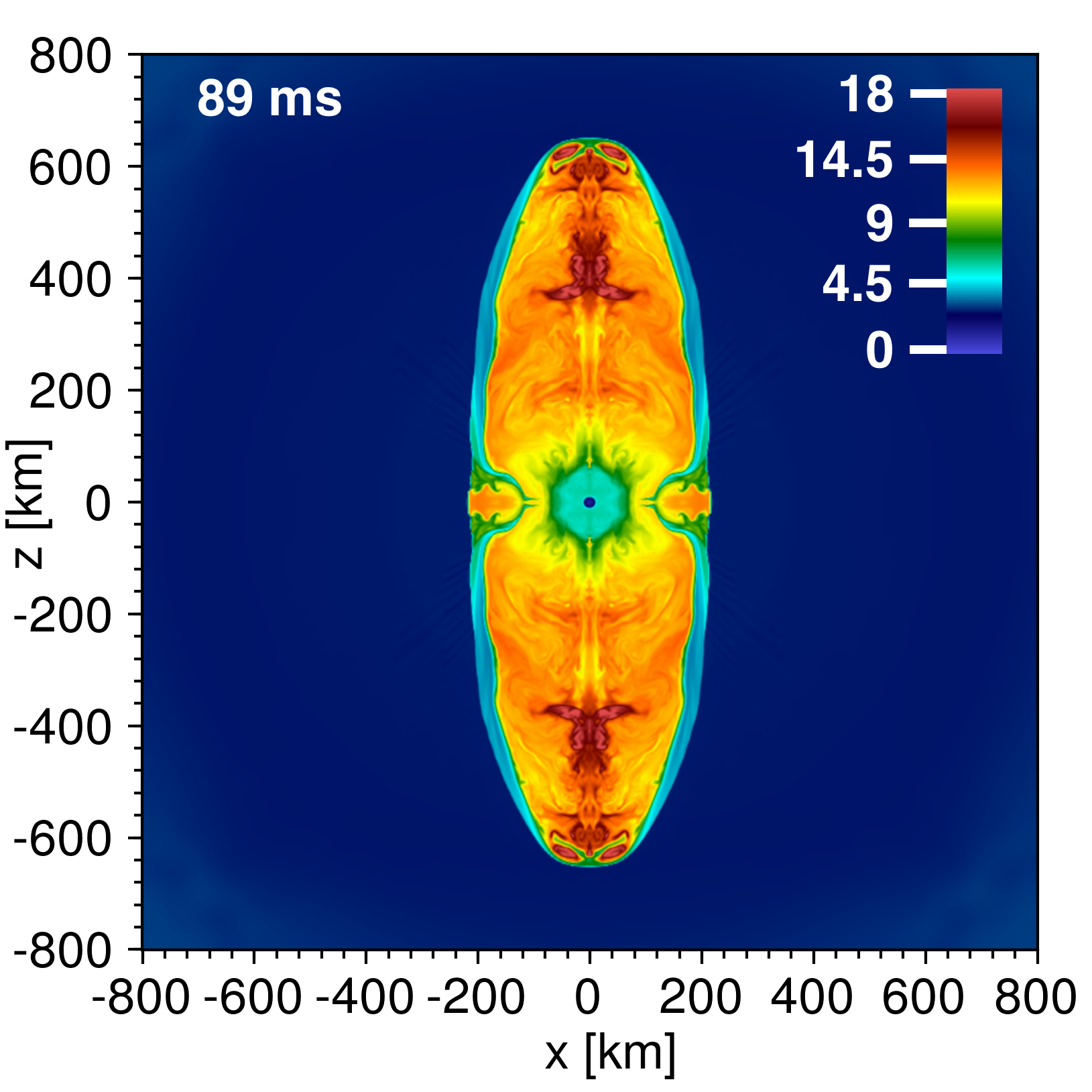}
\includegraphics[width=0.33\textwidth]{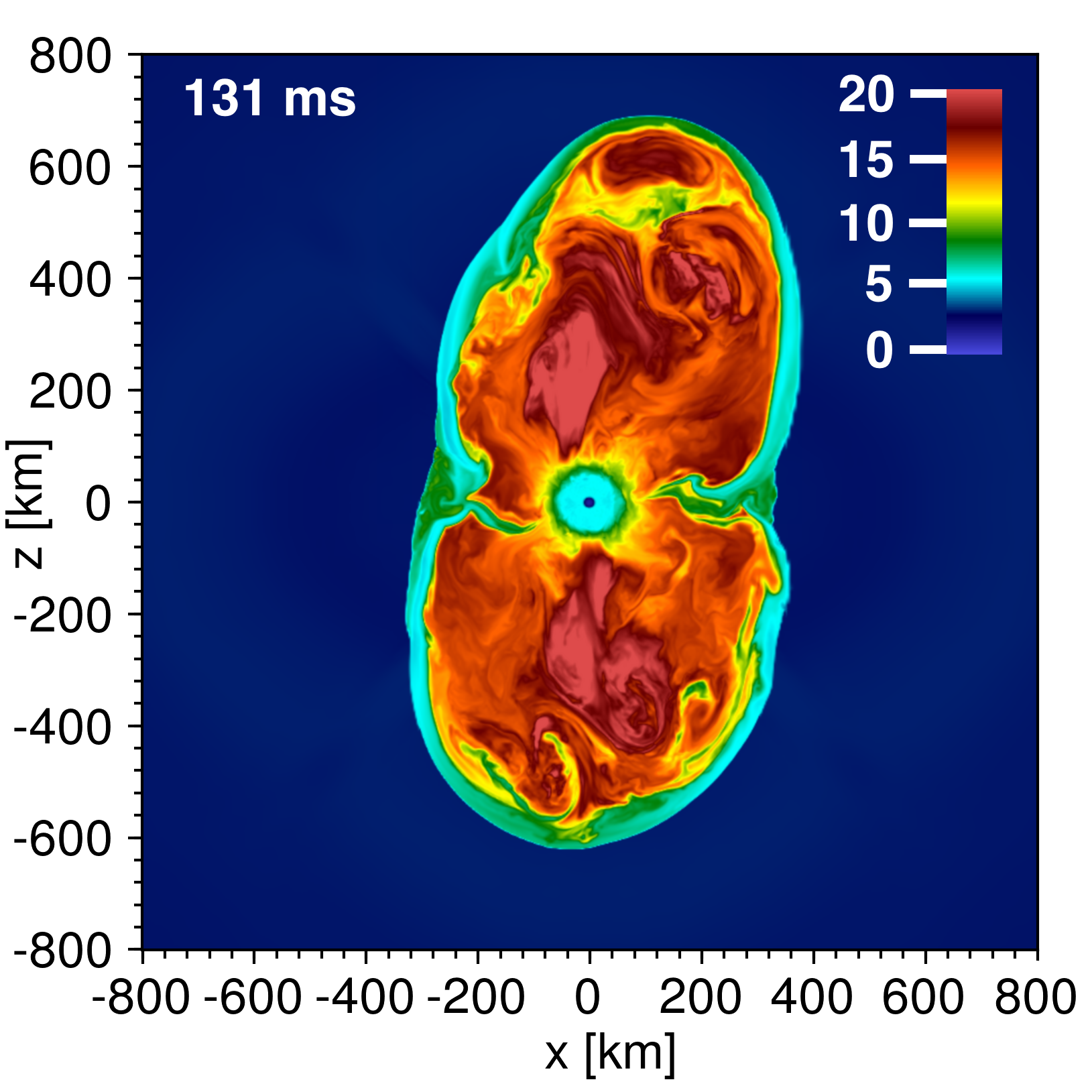}
\vspace{0.1cm}
\caption{Meridional slices ($xz$-plane, $z$ being the vertical) of specific
	entropy $s$ in units of $\mathrm{k_{\mathrm{B}}\, baryon^{-1}}$ for models B13
	(left), B12-sym (center), and B12 (right). The rendering size is
	$1600\, \mathrm{km}\times 1600\, \mathrm{km}$ and times after core
	bounce for model B13, B12-sym, and B12 are $17\, \mathrm{ms}$, $89\,
	\mathrm{ms}$, and $131\, \mathrm{ms}$, respectively. The colormaps vary
	slightly to best capture the dynamics of each simulation and are shown
in the panels. B13 and B12-sym show a clear jet explosion, while B12 explodes
in a dual-lobe fashion due to a kink instability of the
jet~\citep{moesta:14b}.} 
\label{fig:2dsimcmp} 
\vspace{0.5cm} 
\end{figure*}
\begin{figure*}[t]
\includegraphics[width=0.33\textwidth]{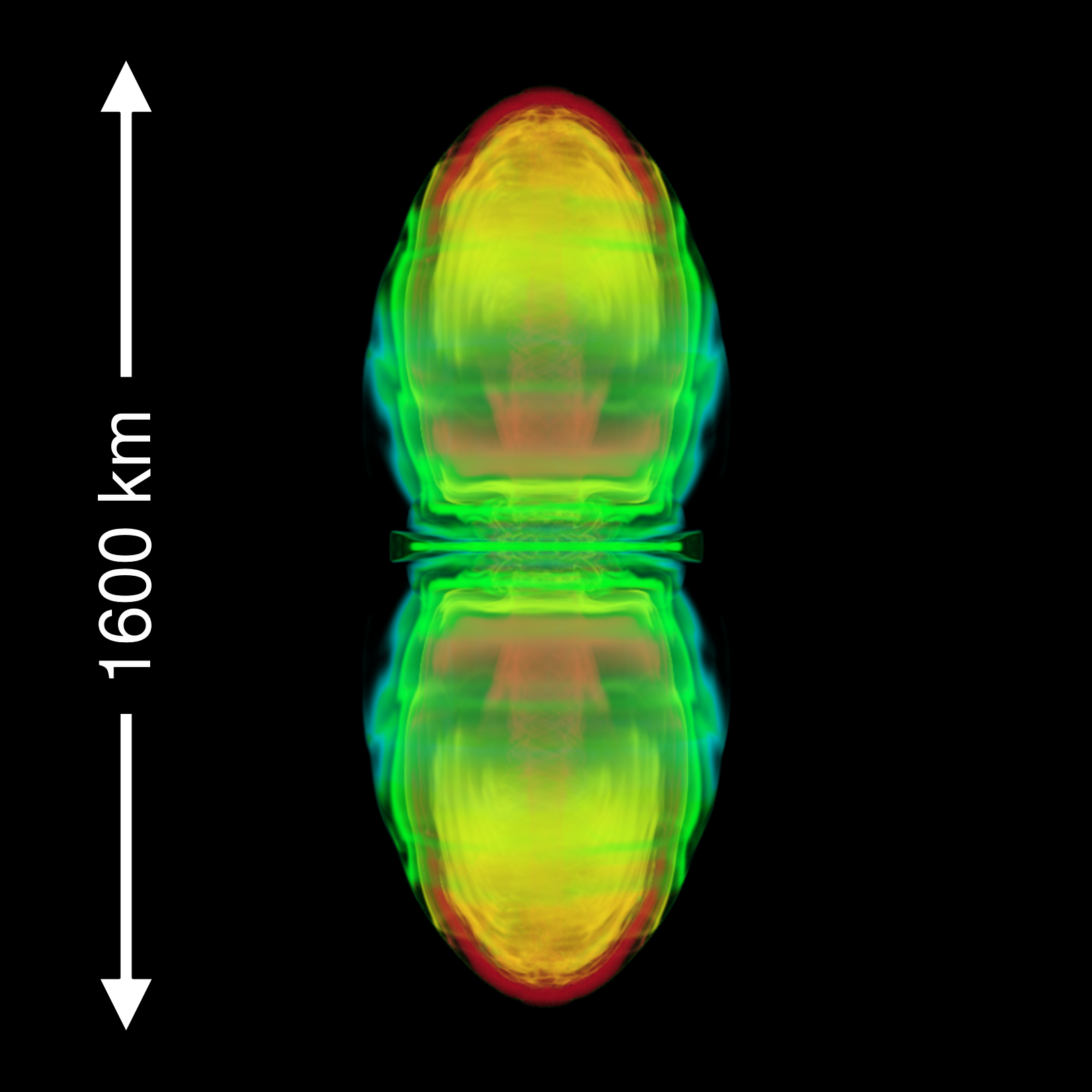}
\includegraphics[width=0.33\textwidth]{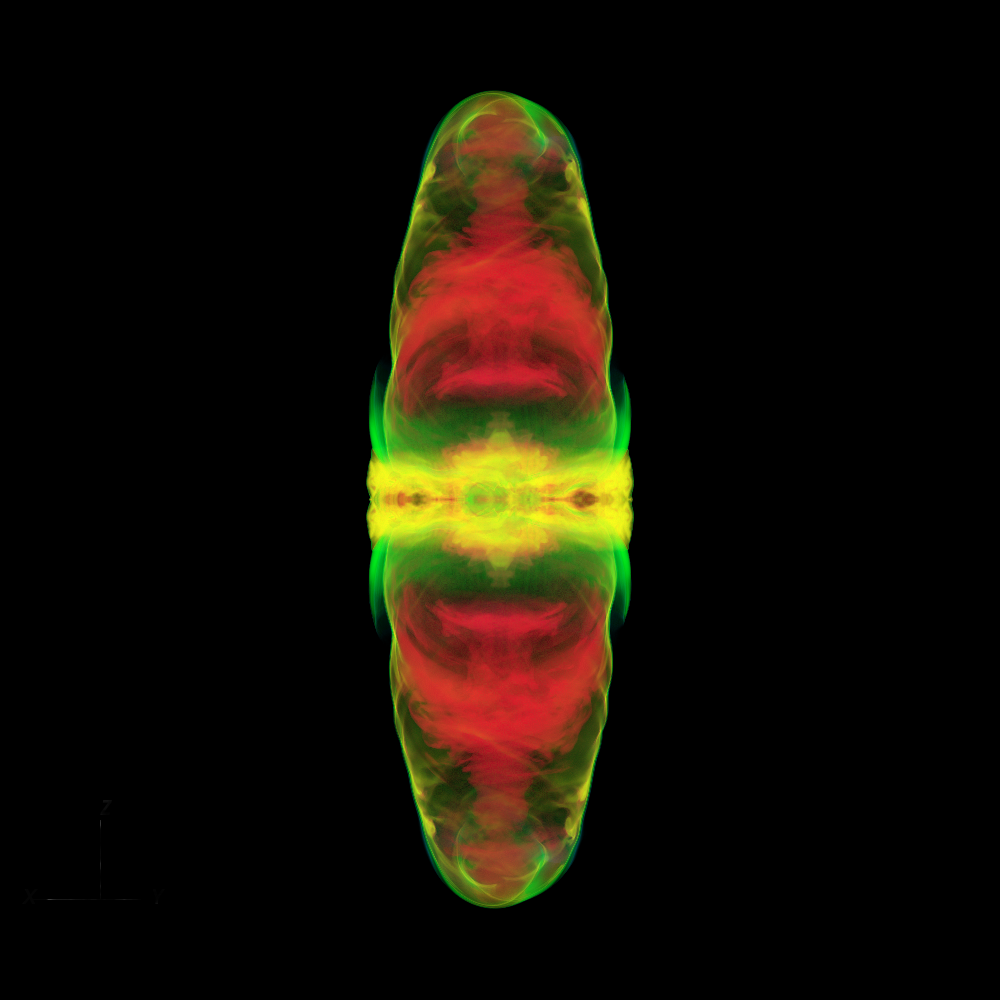}
\includegraphics[width=0.33\textwidth]{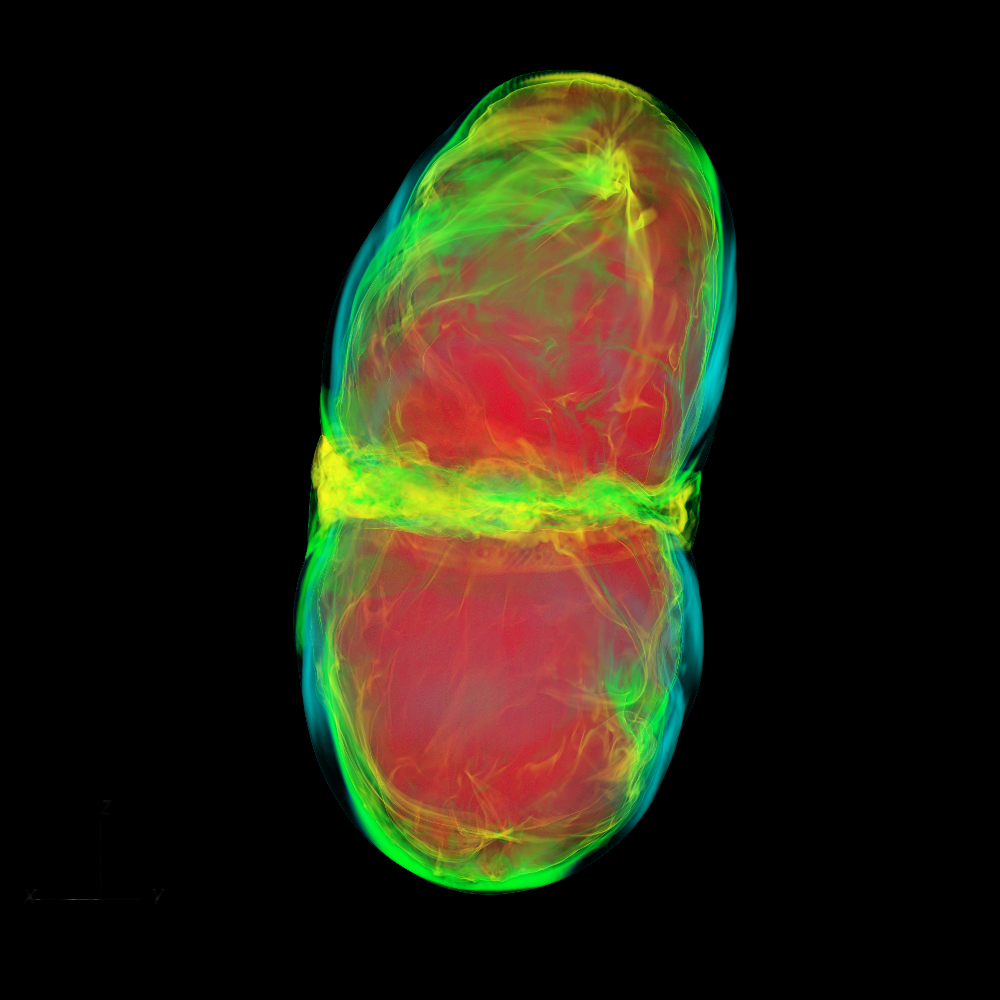}
\vspace{0.1cm}
\caption{Volume renderings of specific entropy for models B13 (left), B12-sym (center), and 
	B12 (right) at the same times as in Fig.~\ref{fig:2dsimcmp}.  The
	z-axis is the rotation axis of the PNS and we show $1600\,
	\mathrm{km}$ on a side. The colormaps vary for the different models
	but are generally chosen such that blue corresponds to lower entropy
	material of $s \simeq 4\, k_{\mathrm{B}}\, \mathrm{baryon}^{-1}$, cyan to
	$s \simeq 5\, k_{\mathrm{B}}\, \mathrm{baryon}^{-1}$ indicating the shock
	surface, green to $s \simeq 6\, k_{\mathrm{B}}\, \mathrm{baryon}^{-1}$, yellow to
	$s \simeq 8 k_{\mathrm{B}}\, \mathrm{baryon}^{-1}$, and red to higher entropy
	material at $s \simeq 10-12\, k_{\mathrm{B}}\,
        \mathrm{baryon}^{-1}$.} 
\label{fig:3dsimcmp} 
\vspace{0.5cm} 
\end{figure*}

Magnetorotational core-collapse
supernovae~\citep{bisno:70,leblanc:70,meier:76,wheeler:02,moesta:14a} are
promising candidate sites for $r$-process
nucleosynthesis~\citep{winteler:12,nishimura:15,nishimura:17}. They are also proposed to be
the engines driving hyperenergetic supernovae from stripped-envelope
progenitors, classified as Type Ic-bl (H/He deficient, broad spectral lines).
The amount of ejected $r$-process material has been found to be $\sim 10^{-3} -
10^{-2}\, M_{\odot}$ in previous studies~\citep{winteler:12,nishimura:15},
similar to what is expected from binary neutron star mergers (e.g.,
\citep{hotokezaka:13}). Magnetorotational supernovae also have the potential to
enrich galaxies early in their cosmic history, as the massive progenitor stars
of these explosions live fast and die young. They therefore offer an intriguing
alternative channel for $r$-process enrichment, especially at low
metallicities.

Magnetorotationally driven supernovae require rapid iron core rotation of the
progenitor star ($P_0 \simeq
\mathcal{O}(1)\,\mathrm{s}$)~\citep{ott:06spin,burrows:07b,moesta:14b} to
form a ms-protoneutron star (PNS) after collapse. 
An additional, magnetar-strength toroidal field then funnels accreted material
into a jet propagating along the rotation axis of the
star~\citep{meier:76,wheeler:02,burrows:07b,moesta:14b,obergaulinger:17a}. This
field can be created by flux-compression from a highly magnetized progenitor
core or via amplification by the \emph{magnetorotational instability}
(MRI,~\citealt{balbus:91,akiyama:03,obergaulinger:09}) and dynamo
action~\citep{moesta:15} in the early postbounce evolution of the protoneutron
star. 

The thermodynamic conditions in the jet-driven outflows typical for these
explosions likely differ from those in neutrino-driven CCSNe.
Material in the outflows is highly-magnetized, underdense, and neutron-rich
(electron fraction $Y_e \simeq 0.1-0.3$). These are ideal conditions for rapid
neutron capture ($r$-process) nucleosynthesis \citep[e.g.,][]{hoffman:97,
meyer:97}. Previously, the $r$-process nucleosynthetic signatures of jet-driven
CCSNe have been studied with axisymmetric (2D) and three-dimensional (3D)
magnetohydrodynamic (MHD) simulations. \cite{winteler:12} found robust
$r$-process nucleosynthesis consistent with the solar abundance pattern for a
3D simulation of the collapse of a highly magnetized progenitor core. The
simulation exhibited a strong jet explosion that was not disrupted by an
$m=1$-kink instability~\citep{moesta:14b}. This was caused by the strong
assumed poloidal field ($B_{\mathrm{pol}} = 5\,\times10^{12}\, \mathrm{G}$),
stabilizing the outflow, since the stability criterion depends on the ratio of
toroidal over poloidal field~\citep{kruskal:58}. The explosion dynamics seen in
their simulations are similar to 2D simulations of jet-driven
CCSNe~\citep{burrows:07b,takiwaki:12}.  \cite{nishimura:15} studied the
$r$-process nucleosynthetic signatures of a range of 2D axisymmetric MHD CCSN
simulations and found that in prompt explosions ($t_{\mathrm{exp}} \leq 50\,
\mathrm{ms}$) a robust $r$-process abundance pattern is recovered, while for
delayed explosions the abundance pattern differs from solar above mass number
$A \sim 130$, which includes the second and third $r$-process peaks.

We present results on $r$-process nucleosynthesis from full 3D
dynamical-spacetime general-relativistic MHD (GRMHD) simulations of rapidly
rotating magnetized CCSNe. We carry out simulations with initial field
strengths of $10^{12}\,\mathrm{G}$ and $10^{13}\,\mathrm{G}$ in full
unconstrained 3D. For the $10^{12}\,\mathrm{G}$ case, we compare results with a
simulation starting from identical initial conditions but that is set up to
remain perfectly axisymmetric in its dynamics. We calculate nucleosynthetic
yields by post-processing Lagrangian tracer particles with the open-source
nuclear reaction network \texttt{SkyNet} \citep{lippuner:17b}. We also
investigate the impact of neutrinos on the nucleosynthetic yields by varying
the uncertain neutrino luminosities from our simulations in the nuclear
reaction network calculation.

Our results for a model with initial poloidal $B$-field of
$10^{12}\,\mathrm{G}$ show that the nucleosynthetic signatures of jet-driven
CCSNe are substantially different when simulated in 2D
versus 3D. In 2D, robust second and third peak $r$-process material is
synthesized in the explosion, while in full 3D, nuclei beyond the second peak
are two orders of magnitude less abundant. Only in a simulation starting with a
$10^{13}\,\mathrm{G}$ poloidal magnetic field (which has dynamics similar to
the simulation of \citealt{winteler:12}), do we find a robust $r$-process pattern
that is consistent with the solar $r$-process residuals. These differences are
driven by differing thermodynamic histories of material ejected in the jet.
For a $10^{12}\,\mathrm{G}$ initial magnetic field, we find that ejected
material reaches lower maximum density before being ejected than in the
simulation with the $10^{13}\, \mathrm{G}$ field. As a result, this material
starts with higher electron fractions as it expands and tries to relax to
$\beta$-equilibrium.  Additionally, the ejected material in the slower jet
explosions experiences more neutrino irradiation, which serves to further
increase the electron fraction. Based on this finding, we conclude
that only jet-driven CCSNe from already strongly magnetized progenitor star
cores ($B \simeq 10^{13}\,\mathrm{G}$) are a viable site for production of the
third $r$-process peak. It is however unrealistic to expect such strongly
magnetized progenitor cores from standard models of massive stellar evolution.

This paper is organized as follows. In Sec.~\ref{sec:methods}, we present the
physical and computational setup and numerical methods used. In
Sec.~\ref{sec:simdynamics}, we present the simulation dynamics, followed by a 
description of the ejecta dynamics in Sec.~\ref{sec:tracerdynamics}. We discuss
the properties of the ejected material in Sec.~\ref{sec:ejecta} before
concluding with a discussion of our findings in Sec.~\ref{sec:discussion}.

\begin{figure}[t]
\includegraphics[width=0.47125\textwidth]{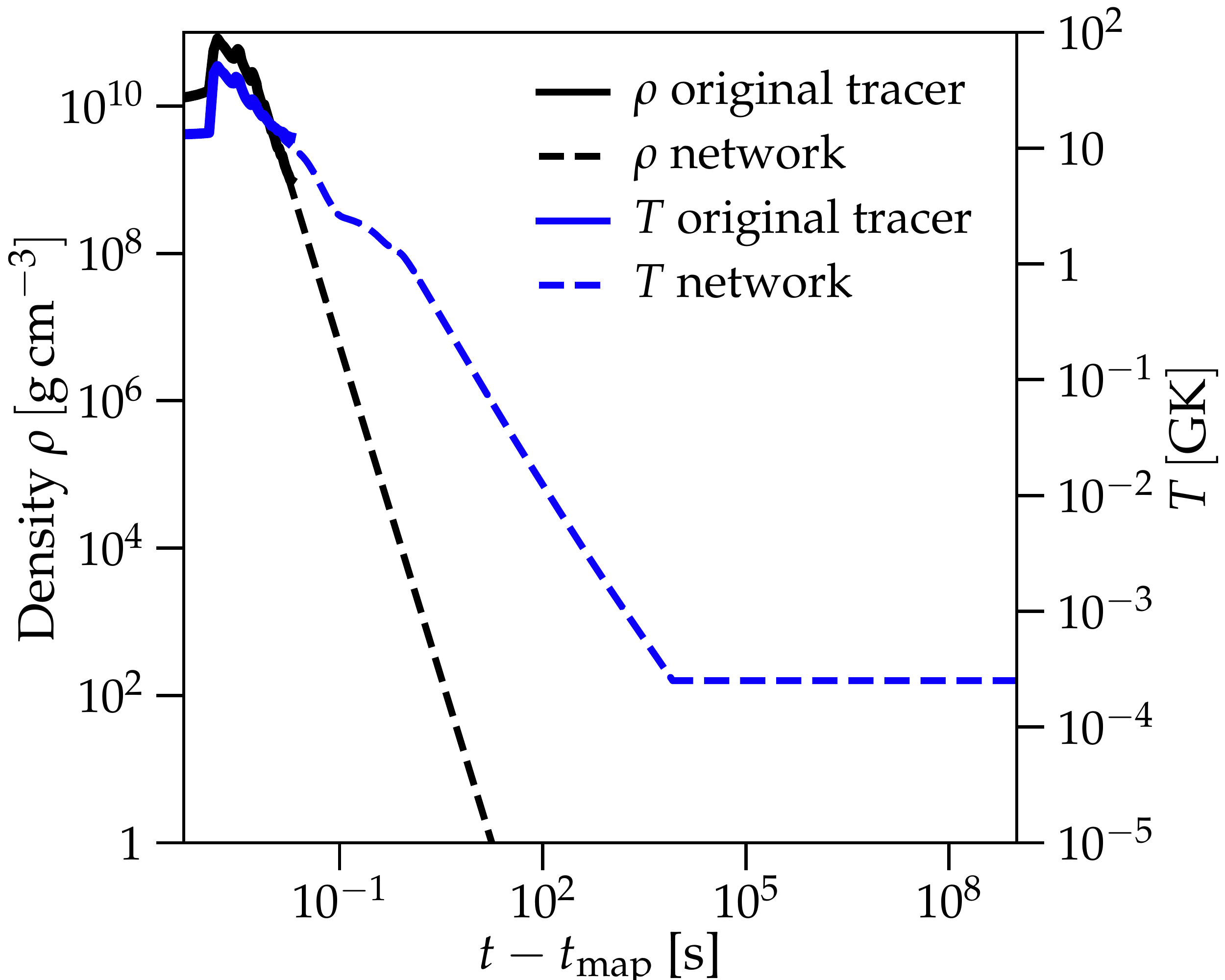}
\caption{Density $\rho$ (black) and temperature (blue) for a representative
	tracer for simulation B13. Solid lines at early times show the original
	tracer particle data, while dashed lines at later times show data
	extrapolated by the network under the assumption of homologous
expansion. The network expands the particle with $\rho(t) = \rho_0 \cdot
t^{-3}$ and $T(t) = T_0 \cdot t^{-1}$ until a minimum temperature is reached.} 
\label{fig:tracerexample} 
\vspace{0.5cm} 
\end{figure}

\section{Methods and Setup}
\label{sec:methods}
\begin{figure*}[t]
\includegraphics[width=0.32\textwidth]{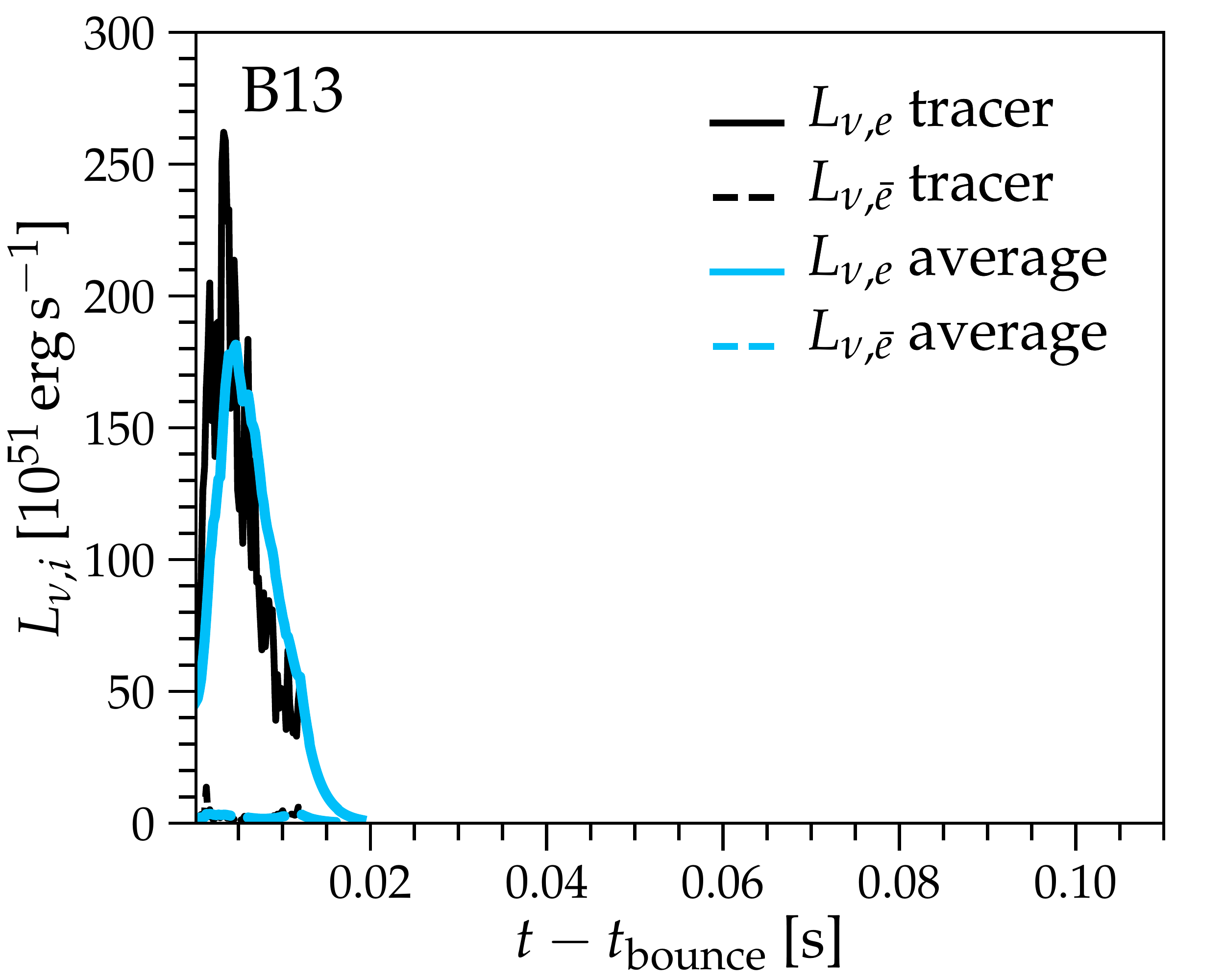}
\includegraphics[width=0.32\textwidth]{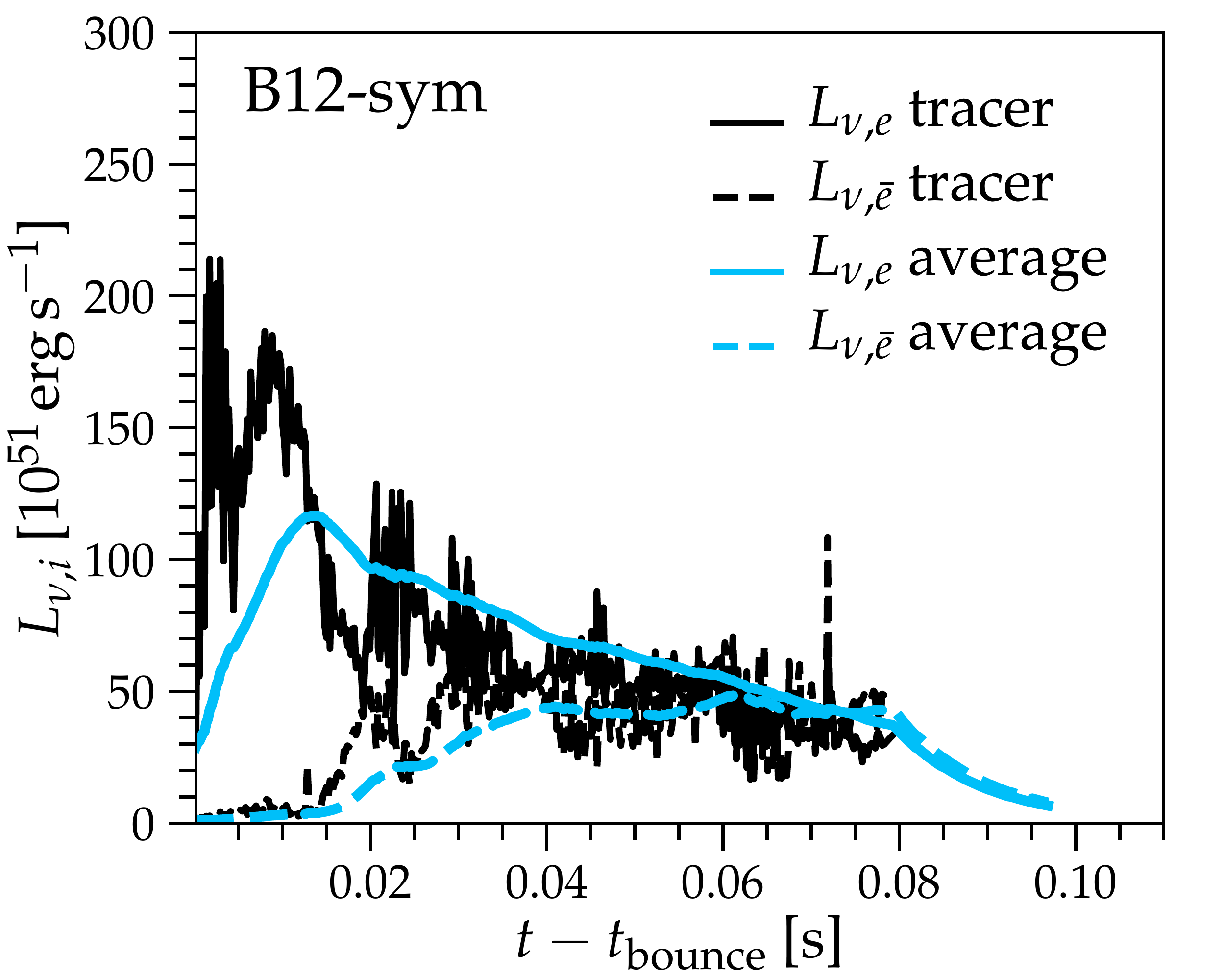}
\includegraphics[width=0.32\textwidth]{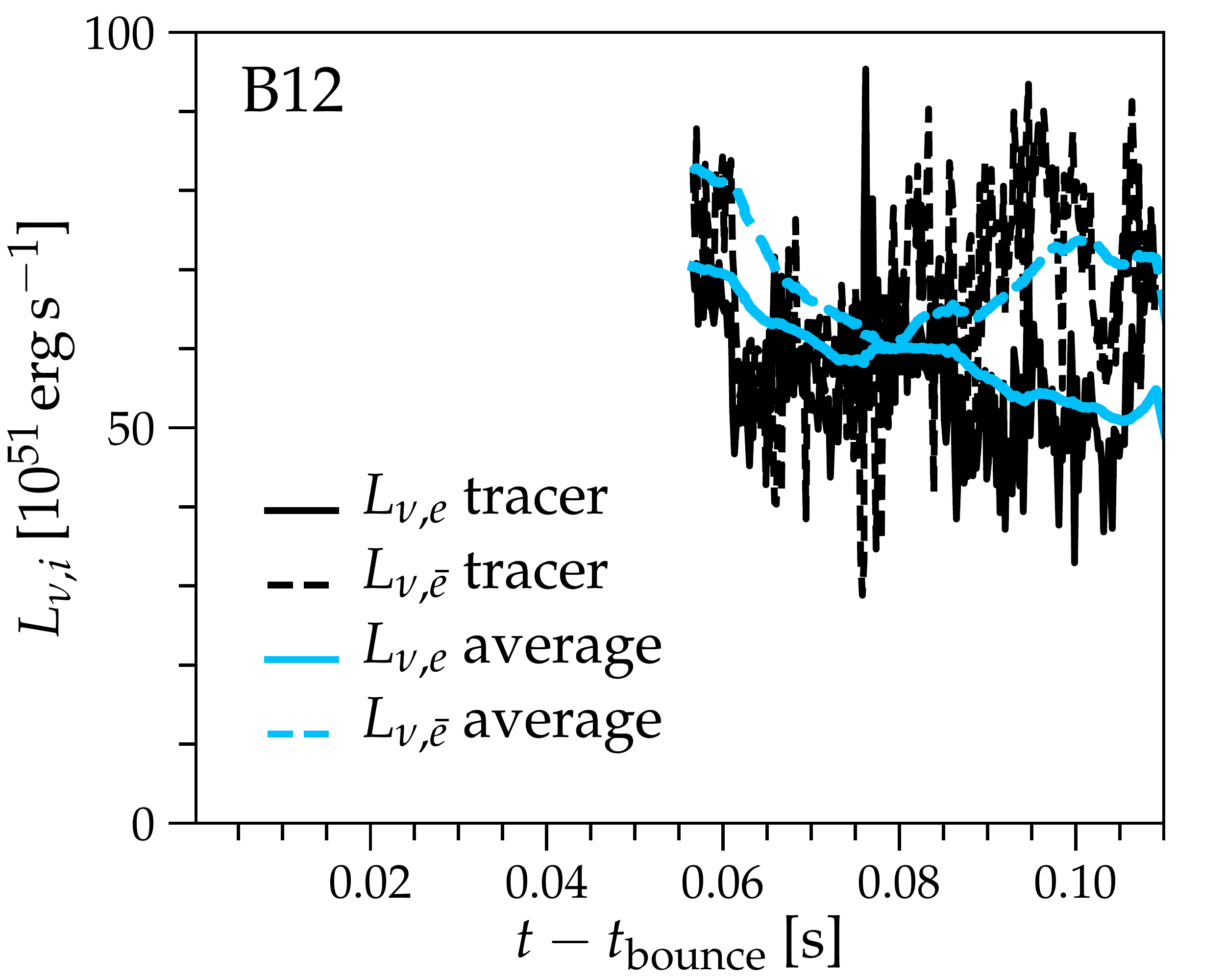}
\caption{Neutrino luminosities for both electron $L_{\nu,e}$ and 
electron anti-neutrinos $L_{\nu,\bar{e}}$ as a function of 
postbounce time for representative single tracer particles from simulation B13
(left), B12-sym (center), and B12 (right). We note that we map the initial
tracer distribution at different times onto the simulation to ensure maximum
control over the number of tracer particles ejected in the outflow. For
simulations B13 and B12-sym, we map the tracer particles at time
$t-t_{\mathrm{bounce}} \simeq 0\, \mathrm{ms}$, for simulation B12 at
$t-t_{\mathrm{bounce}} \simeq 80\, \mathrm{ms}$.} 
\label{fig:lum_cmp} 
\vspace{0.5cm} 
\end{figure*}

\subsection{Simulations}

We employ ideal GRMHD with adaptive mesh refinement (AMR) and spacetime
evolution provided by the open-source \texttt{Einstein Toolkit}
~\citep{moesta:14a,et:12}. GRMHD is implemented in a finite-volume fashion with
WENO5 reconstruction~\citep{reisswig:13a, tchekhovskoy:07} and the HLLE Riemann
solver \citep{HLLE:88} and constrained transport \citep{toth:00} for
maintaining $\mathrm{div} \vec{B} = 0$. We employ the $K_0 = 220\,\mathrm{MeV}$
variant of the equation of state of \cite{lseos:91} and the neutrino
leakage/heating approximations described in \cite{oconnor:10} and
\cite{ott:12a}. At the precollapse stage, we cover the inner
$\sim$$5700\,\mathrm{km}$ of the star with four AMR levels in a Cartesian grid
and add five more during collapse. After bounce, the PNS is
covered with a resolution of $\sim$$370\,\mathrm{m}$ and the AMR grid
structures consists of boxes with extents [$5674.0\,\mathrm{km}$,
$3026.1\,\mathrm{km}$, $2435.1\,\mathrm{km}$, $1560.3\,\mathrm{km}$,
$283.7\,\mathrm{km}$, $212.8\,\mathrm{km}$, $144.8\,\mathrm{km}$,
$59.1\,\mathrm{km}$, $17.7\,\mathrm{km}$]. The coarsest resolution is $h =
94.6\, \mathrm{km}$ and refined meshes differ in resolution by factors of 2. We
use adaptive shock tracking to ensure that the shocked region is always
contained in the mesh refinement box with resolution $h = 1.48\,
\mathrm{km}$.\\\\
 
We draw the $25$-$M_\odot$ (at zero-age-main-sequence) presupernova model E25
from \cite{heger:00}. While this model includes rotation, we parameterize the
initial rotation law to match the simulations in \cite{moesta:14b}. The rotation
law is cylindrical and axisymmetric following \cite{takiwaki:11} equation (1)
and as in \cite{moesta:14b}, 
\begin{equation} 
	\Omega(x,z) = \Omega_0 \frac{x_0^2}{x^2+x_0^2}\, \frac{z_0^4}{z^4+z_0^4}, 
\end{equation} 
with an initial central angular velocity $\Omega_0 = 2.8\, \mathrm{rad}\,
\mathrm{s}^{-1}$. The fall-off in cylindrical radius and vertical position is
controlled by parameters $x_0 = 500\,\mathrm{km}$ and $z_0 =
2000\,\mathrm{km}$, respectively. The rotation parameter of our setup is
$\beta_{\mathrm{rot}} = 0.1\, \%$ where $\beta_{\mathrm{rot}} \equiv T/|W|$ is the
ratio of kinetic and gravitational potential energy. This is consistent with
typical GRB-oriented progenitors models (i.e., E25 in \cite{heger:00} has
$\beta_{\mathrm{rot}} \sim 0.15\, \%$). We set up the initial magnetic field
using a vector potential of the form $$A_r = A_\theta = 0; A_\phi = B_0
({r_0^3})({r^3+r_0^3})^{-1}\, r \sin \theta,$$ where $B_0$ controls the
strength of the field. In this way, we obtain a modified dipolar field
structure that stays nearly uniform in strength within radius $r_0$ and falls
off like a dipole at larger radii. We choose $r_0 = 1000\, \mathrm{km}$ to
match the initial conditions of model B12X5$\beta$0.1 of the 2D study of
\cite{takiwaki:11} and the 3D study of \cite{moesta:14b}. We perform
simulations for two different initial magnetic field strengths, $B_{0} =
10^{13}\, \mathrm{G}$ (B13 from here on) and $B_{0} = 10^{12}\, \mathrm{G}$
(B12).

\begin{table} \centering \begin{threeparttable} \caption{Initial magnetic field
	strength and perturbation setup (in velocity) for the three simulations
considered here.} \begin{tabular}{lccc} \hline \hline Simulation & B13 &
	B12-sym & B12\\ \hline $B_{\mathrm{pol}}$ [G] & $10^{13}$ & $10^{12}$ &
	$10^{12}$\\ Perturbations & None & None & $0.01 \times
	\left|\vec{v}\right|$  \\ \hline
\end{tabular}
\label{tab:simoverview}
\vspace{0.2cm}
\end{threeparttable}
\end{table}

We perform simulations in full, unconstrained 3D. For model B12, we add random
perturbations with magnitude $1\%$ of the velocity at the start of the
simulation. In model B12-sym we do not add perturbations and the simulation
therefore evolves identical to an octant symmetry 3D (90-degree rotational
symmetry in the $x-y$ plane and reflection symmetry across the $x-y$ plane)
simulation. In this way, we reproduce the dynamics of an axisymmetric simulation while
keeping the tracer particle setup and distribution identical between the simulations. In
model B13, the ten times stronger initial poloidal magnetic field prevents the
disruption of the jet by a kink-instability~\citep{moesta:14b} as the key
quantity for instability is the ratio of toroidal over poloidal field (see
Eq.~(2) in \citealt{moesta:14b}). As a result, simulations with and without
perturbations for model B13 are nearly identical and we only present results
for a simulation without added perturbations. Model B13 is closest in dynamics
to the model presented in \cite{winteler:12}, while model B12-sym mimics the
dynamics of the prompt (explosion within 50 ms after core bounce) 2D jet
explosions in \cite{takiwaki:11} and \cite{nishimura:15}. We summarize the
initial magnetic field strengths and perturbation setups used in the
simulations in Table~\ref{tab:simoverview}.

\subsection{Tracer particles and postprocessing}

We extract the thermodynamic conditions of ejected material
using Lagrangian tracer particles. We place $10^5$ tracer particles on each of
the simulations. We limit particles to $ 30\, \mathrm{km} \leq r \leq 1000\,
\mathrm{km}$ for all simulations to ensure high enough resolution in particle
mass but also guarantee that the infall time for the outermost shell of
particles is longer than the simulated time. The tracers are uniformly spaced,
so that they represent regions of constant volume. Each tracer particle gets
assigned a mass taking into account the density at its location and the volume
the particle covers. Tracer particles are advected passively with the fluid
flow and data from the 3D simulation grid are interpolated to the tracer
particle positions. In this way, we record the thermodynamic conditions and
neutrino luminosities the particles encounter as a function of time. To
calculate the ejected mass in the explosion we only take dynamically unbound
particles into account. We determine whether a particle is unbound 
by determining the total specific energy is positive (and define the specific
energy as the sum of internal, kinetic, and magnetic energy). 

For models B13 and B12-sym, we map the tracer particle
distribution onto the simulation shortly after core bounce as both of these
models explode within the first 40 ms of postbounce evolution. Model B12 takes
considerably longer to explode and we map the tracer distribution onto the
simulation shortly before transition to explosion at time
$t-t_{\mathrm{bounce}} = 80\, \mathrm{ms}$. This allows us to ensure that we
have a sufficient number of tracer particles in the outflows along the rotation
axis of the core.
We postprocess the particles with the open-source nuclear reaction network
\texttt{SkyNet} of \cite{lippuner:17b}. The network includes 7843 isotopes up to
isotope $^{337}$Cn. Forward strong rates are taken from the JINA REACLIB
database~\citep{cyburt:10}, and inverse rates are computed assuming detailed
balance. Weak rates are taken from \cite{fuller:82}, \cite{oda:94},
\cite{langanke:00}, or otherwise from REACLIB. REACLIB also provides nuclear
masses and partition functions. \texttt{SkyNet} evolves the temperature via the
computation of source terms due to the individual nuclear reactions and
neutrino interactions.

Computations for each particle start from nuclear statistical equilibrium
(NSE). We start the network as soon as the temperature drops below $T=25\,
\mathrm{GK}$. The initial conditions for the network calculation are taken at
this time. The neutrino luminosity data from the particle trajectories are noisy
due to interpolation effects and the very high time resolution at which the tracer
particles record the neutrino luminosities. We therefore compute a
moving-window time average of the neutrino luminosities as 
$\nu_{\mathrm{av},i} = \alpha\cdot\bar{\nu}_{i} +
(1.0-\alpha)\cdot\nu_{\mathrm{av},i-1}$,where $i$ denotes the current timestep
data and $i-1$ the previous one. We choose a weight function for each dataset
in the moving average as $\alpha = 2\cdot(n+1.0)^{-1}$, with $n = 40$ and keep the
neutrino luminosities constant after the end of the particle data. In cases
in which the particle data in the simulation does not reach temperatures low
enough for the network calculation to start, we extrapolate the particle data
assuming homologous expansion. We carry out the network calculations to
$10^9\,\mathrm{s}$, which is sufficient to generate stable abundance patterns
as a function of mass number $A$.

\begin{figure*}[t]
\centering
\includegraphics[width=0.32\textwidth]{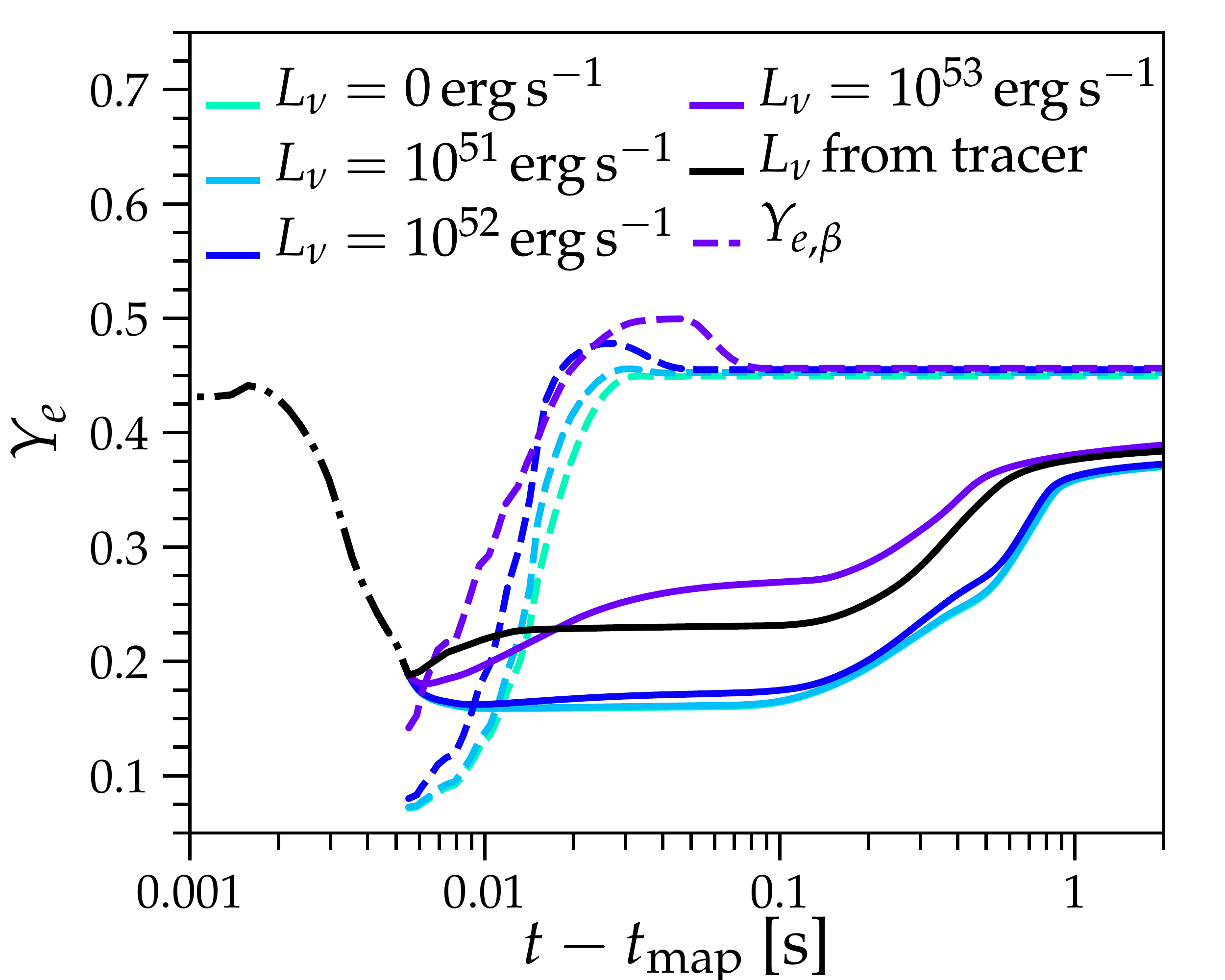}
\includegraphics[width=0.32\textwidth]{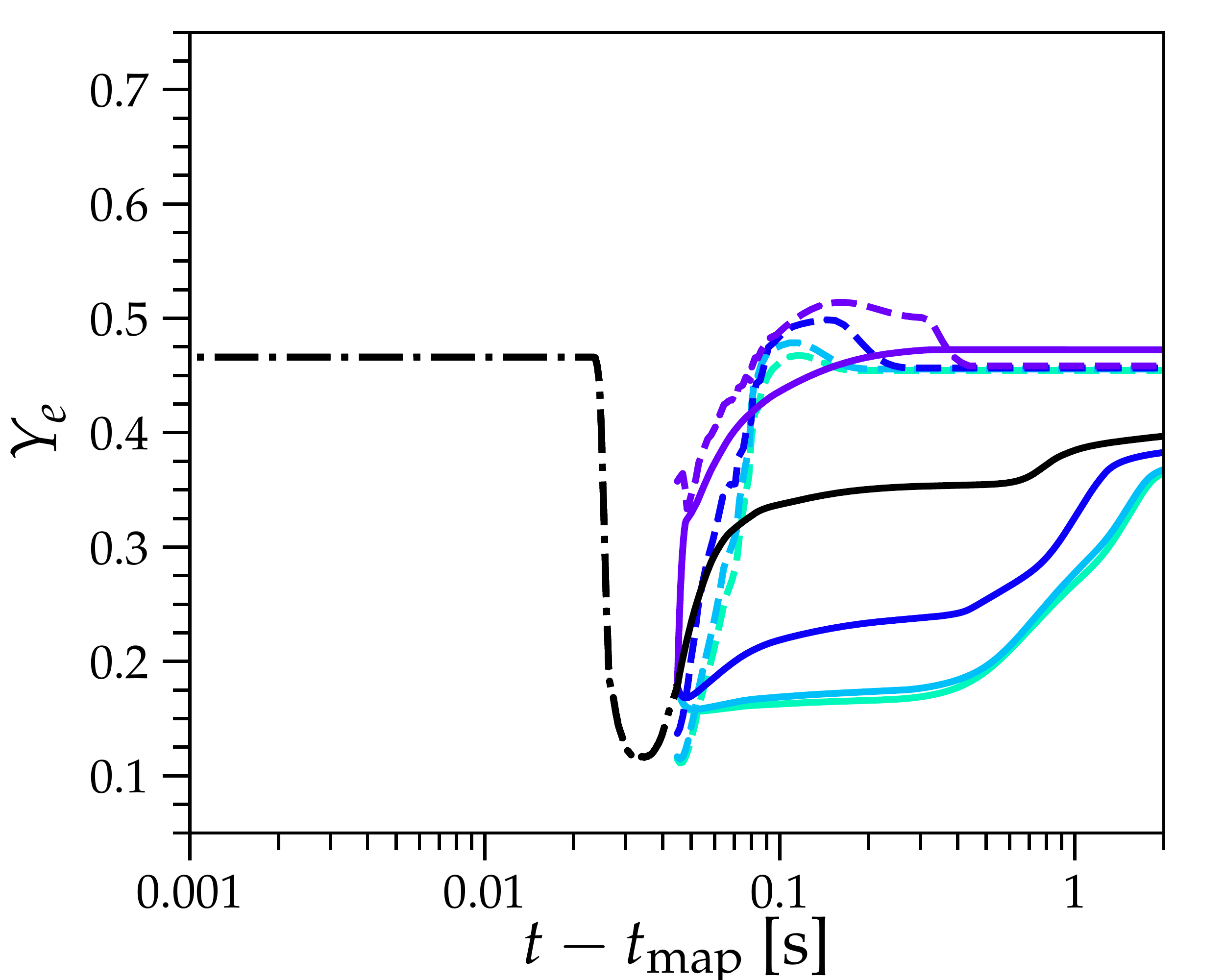}
\includegraphics[width=0.32\textwidth]{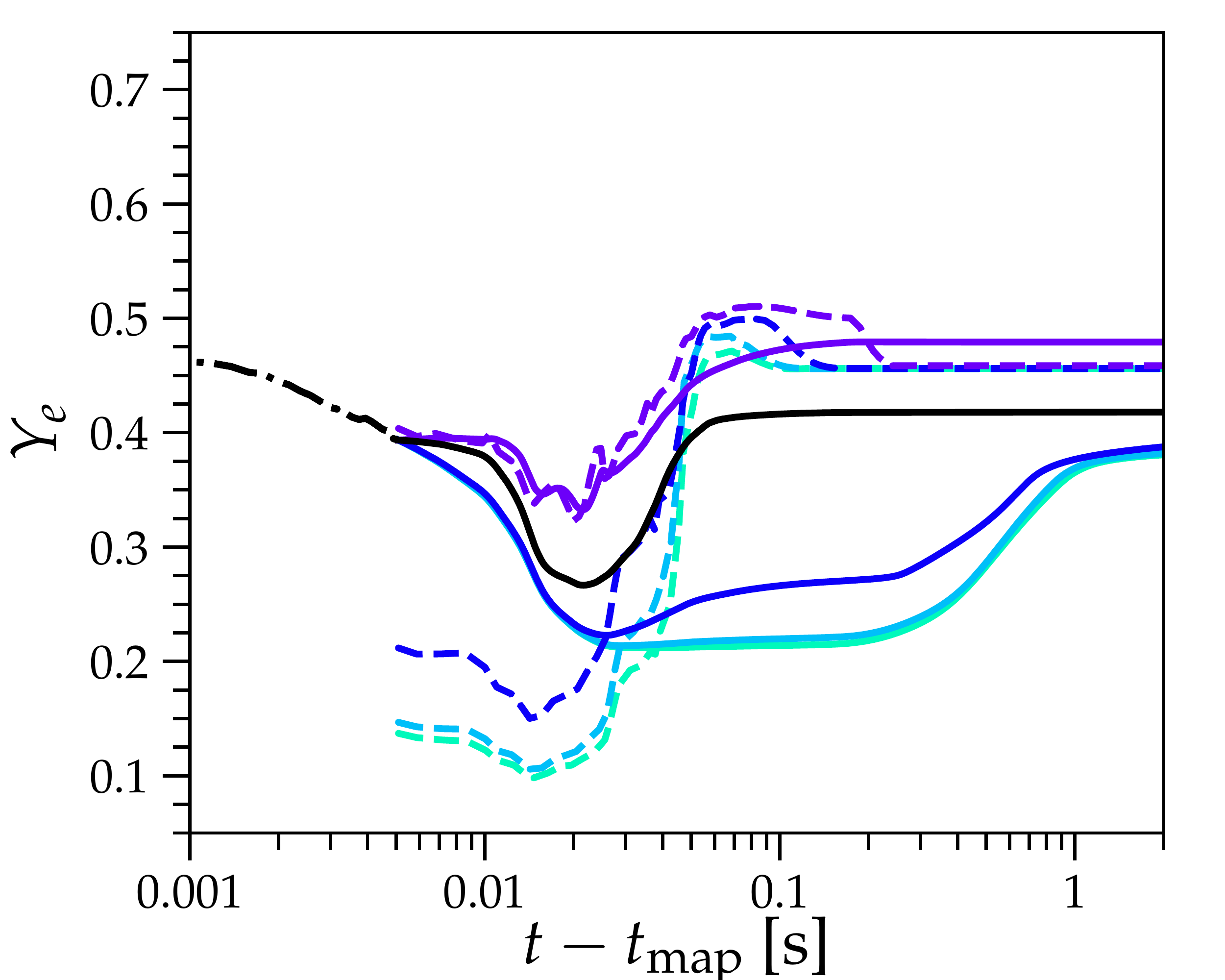}
\includegraphics[width=0.32\textwidth]{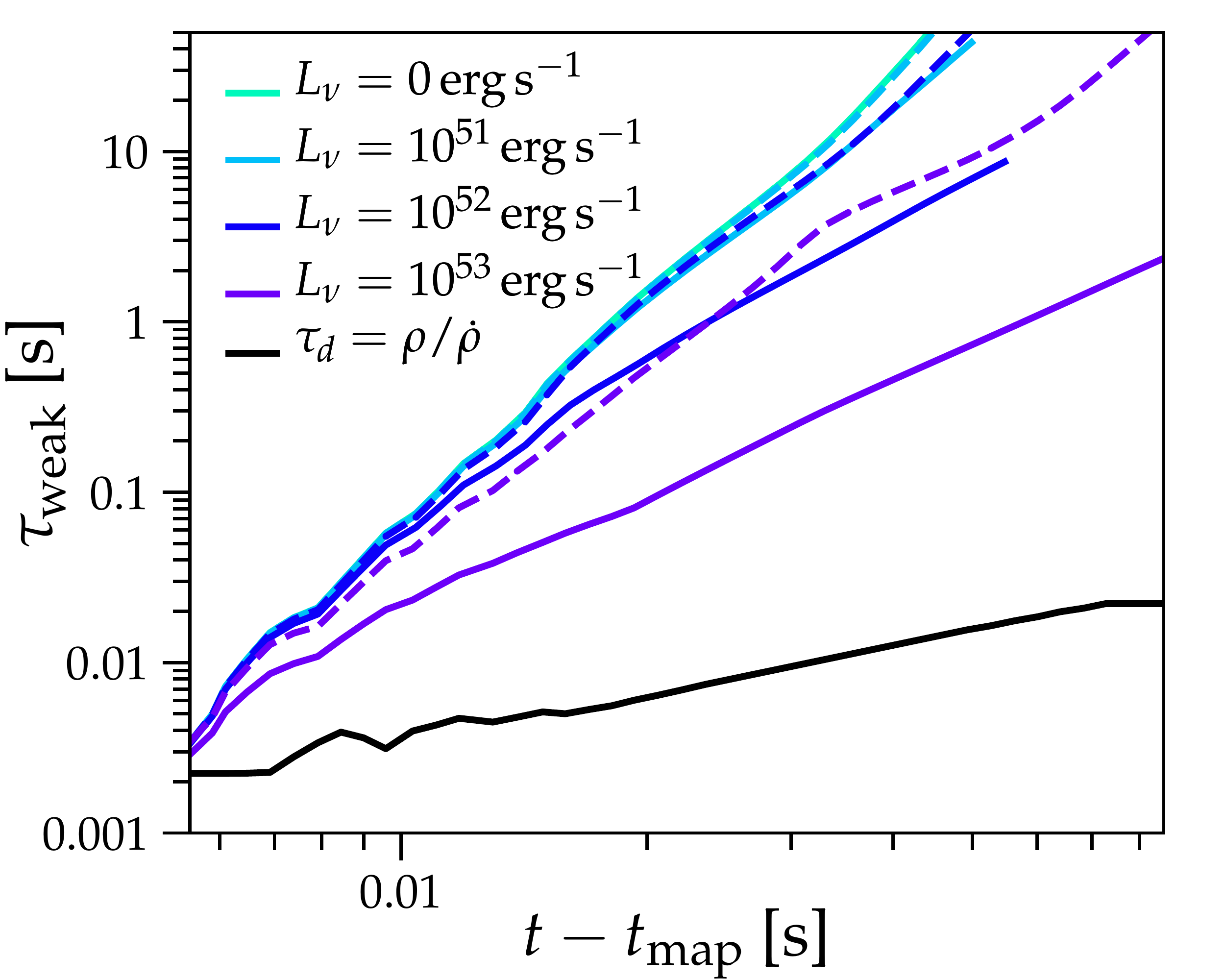}
\includegraphics[width=0.32\textwidth]{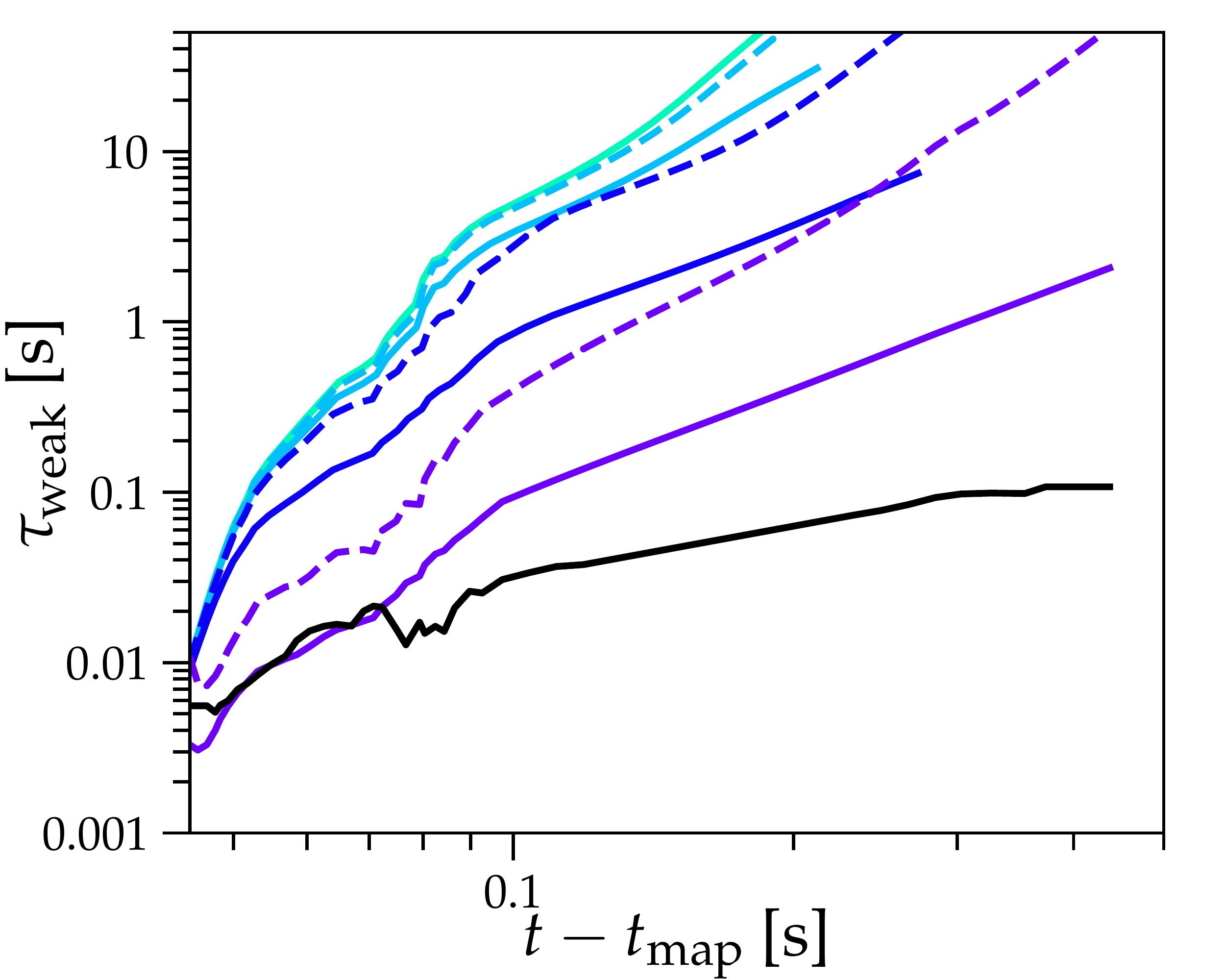}
\includegraphics[width=0.32\textwidth]{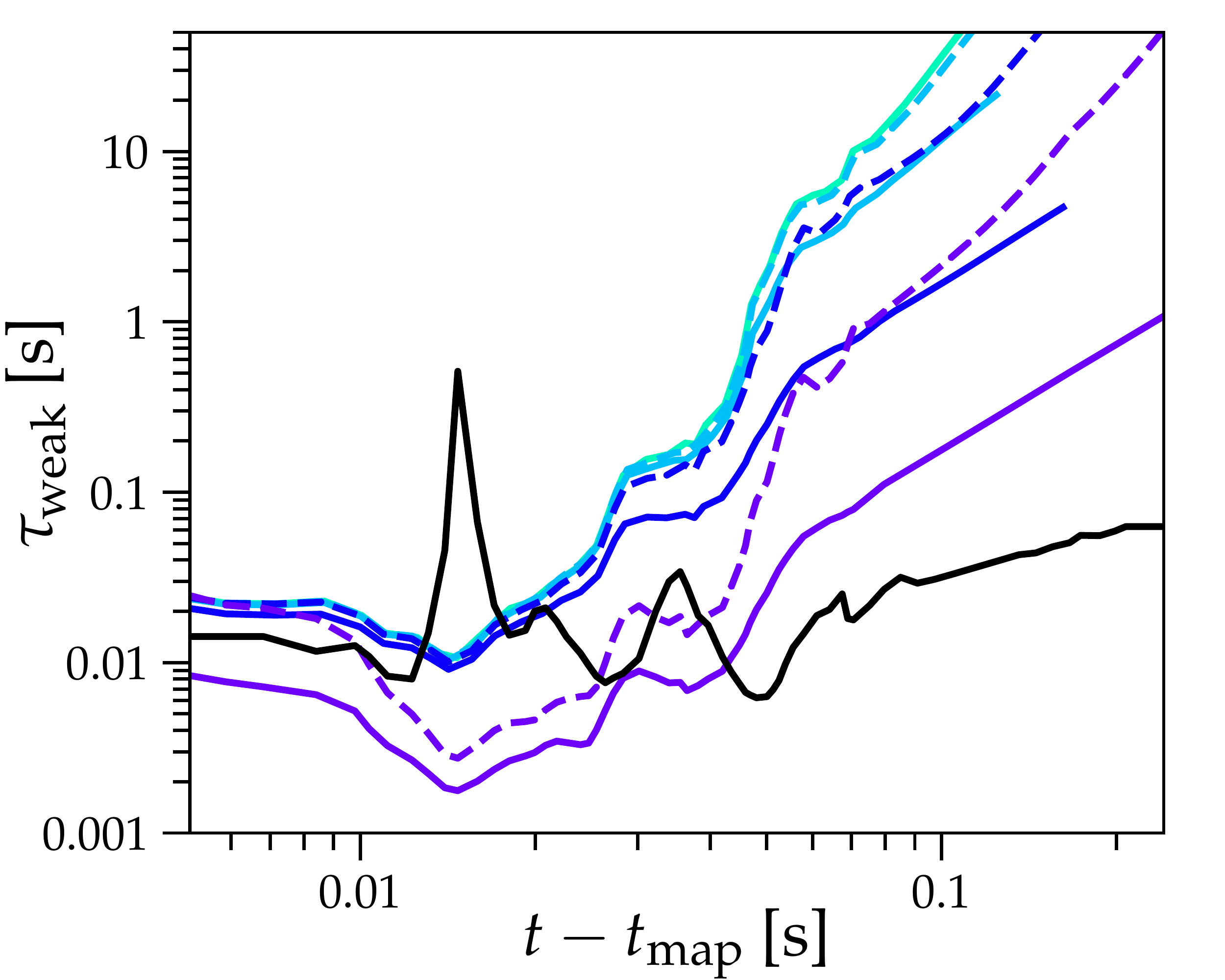}
\caption{Top Row: Electron fraction $Y_e$ as a function of time after mapping
	the particles onto simulation B13 (left), B12-sym (center), and B12
	(right) for representative particles. Differently colored lines
	indicate results for different neutrino luminosities (assuming $L_{\nu}
	= L_{\nu_e} = L_{\nu_{\bar{e}}}$) used in the nuclear reaction network
	calculation. Black lines indicate results using the neutrino
	luminosities from the tracer particles advected with the simulations.
	The dashed lines indicate the evolution of $Y_{e,\beta}$ for each of
	the fixed neutrino luminosity simulations. The particle in simulation
	B13 reaches the lowest $Y_e$ values while the particles in simulations
	B12-sym and B12 turn around at increasing minimum $Y_e$ values. The dot
	dashed lines show the evolution of $Y_e$ in the tracer particles before
the nuclear reaction network calculations begin. Bottom Row: Weak interaction
and dynamical timescales for the same three models. The dashed lines indicate
the lepton capture timescale, $(\lambda_{e^-}+\lambda_{e^+})^{-1}$.}
\label{fig:ye_cmp} 
\vspace{0.5cm}
\end{figure*}

\section{Results}
\label{sec:results}
\subsection{MHD dynamics}

\label{sec:simdynamics}
Collapse and early postbounce evolution proceed identically in B12 and
B12-sym. Core bounce occurs $\sim$$350\,\mathrm{ms}$ after the onset of
collapse for model B12 and $\sim$$450\,\mathrm{ms}$ for model B13. The delay
in core bounce for model B13 is due to the additional support by the 100
times higher magnetic pressure. Shortly after core-bounce the poloidal and toroidal
B-field components reach $B_{\mathrm{pol}},B_{\mathrm{tor}} \sim
10^{15}\, \mathrm{G}$ for model B12 and $B_{\mathrm{pol}},B_{\mathrm{tor}} \sim
10^{16}\, \mathrm{G}$ for B13.

In model B13, the hydrodynamic shock launched at bounce, still approximately
spherical, never stalls, but continues to propagate into a jet explosion along
the rotation axis (left panels of Fig.~\ref{fig:2dsimcmp} and
Fig.~\ref{fig:3dsimcmp}). The jet is powered by the extra pressure and stress
from the strong magnetic field. The shock propagates at mildly relativistic
speeds ($v_{\mathrm{jet}} \sim 0.1-0.2\,c$) and reaches $1000 \, \mathrm{km}$ at around
$35\,\mathrm{ms}$ after core bounce. The jet is stabilized against the MHD kink
instability by its large poloidal magnetic field (see stability condition
Eq.~(2) in \citealt{moesta:14b}). A mild $m=0$-deformation is visible in the
outflow in Fig.~\ref{fig:3dsimcmp}.

In models B12 and B12-sym, the bounce shock stalls after
$\sim$$10\,\mathrm{ms}$ at a radius of $\sim$$110\,\mathrm{km}$. At this time,
there is strong differential rotation in the region between the protoneutron
star core and the shock. This differential rotation powers rotational winding
of the magnetic field and amplifies its toroidal component to
$10^{16}\,\textrm{G}$ near the rotation axis within $20\,\textrm{ms}$ of
bounce. The strong polar magnetic pressure gradient, in combination with hoop
stresses exerted by the toroidal field, then launches a bipolar outflow. As in
\cite{moesta:14b}, B12-sym now continues into a jet explosion and reaches
$\sim$$900\,\mathrm{km}$ after $\sim$$100\,\mathrm{ms}$. The expansion speed at
that point is mildly relativistic ($v_{\mathrm{jet}} \simeq 0.1-0.15\, c$). 

The 3D simulation with perturbations B12 starts to diverge from its symmetric
counterpart B12-sym around $\sim$15$\,\mathrm{ms}$ after bounce due to the
non-axisymmetric spiral MHD kink instability \citep{moesta:14b}. The subsequent
3D evolution is fundamentally different from both the effectively axisymmetric
B12-sym model and model B13. The jet is strongly disrupted by the kink
instability, which causes the outflow to cover a larger solid angle. The shock
also propagates at a lower velocity than in simulations B13 and B12-sym
($v_{\mathrm{jet}} \simeq 0.03-0.05 c$).

The developed jet structures of models B13, B12-sym, and B12 are depicted in
Fig.~\ref{fig:2dsimcmp} (meridional slices) and Fig.~\ref{fig:3dsimcmp} (volume
renderings). In both B13 (left) and B12-sym (center), a clean jet emerges, and
propagates at mildly relativistic speeds into the outer layers of the core and
star. For simulation B12 the explosion propagates in dual-lobe fashion, as in
\cite{moesta:14b}, at non-relativistic speeds.  Material in the outflows of
all three simulations is highly magnetized ($\beta = P_\mathrm{gas} /
P_\mathrm{mag} \ll 1$), ranges between $10\, k_{\mathrm{B}}\,
	\mathrm{baryon^{-1}}\leq s \leq 20\, k_{\mathrm{B}}\,
	\mathrm{baryon^{-1}}$ in specific entropy, and is neutron-rich ($0.1
	\leq Y_e \leq 0.4$).

\subsection{Ejecta dynamics}
\label{sec:tracerdynamics}

Material that gets accreted across the shock into the postshock flow is
pushed to higher densities and temperatures as it is advected toward the
PNS. Eventually, some of this infalling material is entrained in
the outflow and ejected. We show the evolution of temperature and density for a
typical tracer particle trajectory from simulation B13 in
Fig.~\ref{fig:tracerexample}. Additionally, we show neutrino luminosities for
both electron and electron antineutrinos in Fig.~\ref{fig:lum_cmp} recorded by
three representative particles from simulations B13, B12-sym, and B12. After the
initial neutronization burst visible for simulations B13 and B12-sym in the
first 20 ms of postbounce evolution, the neutrino luminosities for both
electron neutrinos and antielectron neutrinos converge toward values of $\simeq
5\times 10^{52}\mathrm{erg}\mathrm{s}^{-1}$ and stay approximately constant for
the duration of the simulations. 

Material that is ejected often reaches conditions where weak reactions proceed
rapidly enough for weak equilibrium (or $\beta$-equilibrium) to nearly take
hold. In weak or $\beta$-equilibrium the rate of neutron destruction balances
the rate of proton destruction. For a closed, thermalized system,
$\beta$-equilibrium is characterized by the condition $\mu_{\nu_e} + \mu_n =
\mu_e + \mu_p$, where $\mu_i$ are the chemical potentials of electron
neutrinos, neutrons, electrons, and protons, respectively. For a fixed lepton
fraction $Y_L=Y_e+Y_{\nu_e}$, the condition of $\beta$-equilibrium determines
$Y_e$ and the net electron neutrino fraction $Y_{\nu_e}$. When neutrinos are
not trapped, the material moves toward dynamic $\beta$-equilibrium, which is
not determined by chemical potential equality but rather by rate balance
\citep[e.g,][]{arcones:10}. Considering only captures on neutrons and protons,
dynamic $\beta$-equilibrium is given by the condition \begin{equation} \dot Y_e
= [\lambda_{e^+}(\rho, T, Y_e) + \lambda_{\bar \nu_e}] Y_n -
[\lambda_{e^-}(\rho, T, Y_e) + \lambda_{\nu_e}] Y_p = 0, \end{equation} where
the free proton and neutron fractions are set by NSE and $\lambda_{e-}$,
$\lambda_{e+}$, $\lambda_{\nu_e}$, and $\lambda_{\bar \nu_e}$ are the rates of
electron, positron, electron neutrino, and electron antineutrino capture,
respectively. Eq. (2) is an implicit equation for $Y_e$ in $\beta$-equilibrium
$Y_{e,\beta} = Y_{e,\beta}(\rho, T)$. When material is out of
$\beta$-equilibrium, weak interactions will push its $Y_e$ toward $Y_{e,\beta}$
on a timescale given by $\tau_{weak} = (\lambda_{e^-}(\rho, T, Y_e) +
\lambda_{e^+}(\rho, T, Y_e) + \lambda_{\nu_e} + \lambda_{\bar \nu_e})^{-1}$. If
$\tau_{weak}$ is shorter than the dynamical timescale, $\tau_d = \rho/\dot
\rho$, then the material should relax to a composition determined by $Y_{e,
\beta}$.

The value of $Y_{e, \beta}$ depends both on the imposed neutrino fluxes and the
thermodynamic state of the material, which determines the lepton capture rates.
Generally, the electron capture rate dominates at high densities where electrons
are degenerate, which pushes $Y_{e,\beta}$ to values less than $\sim
0.3$. For fixed entropy at densities between $10^9 - 10^{12} \, \textrm{g
cm}^{-3}$, the $\beta$-equilibrium electron fraction goes down with increasing
density. Therefore, rapidly ejected material that has reached a higher density
will often have a lower $Y_e$ at the time $r$-process nucleosynthesis begins
because it experienced more electron captures. On the other hand, when neutrino
captures dominate (and when there are no nuclei present), one finds
$Y_{e,\beta} \approx \lambda_{\nu_e}(\lambda_{\bar \nu_e} +
\lambda_{\nu_e})^{-1}$ \citep{qian:96}. Since the neutrino emission in CCSNe is
fairly similar in all flavors, $\beta$-equilibrium driven by neutrino captures
generally predicts $Y_{e,\beta}>0.4$. Material ejected in the jet goes through
density regimes where the electron captures dominate and then later through
regimes where the neutrino captures dominate. In many cases, the material is
never able to fully attain $\beta$-equilibrium. 

In the jet-driven SNe considered here, weak reactions at high density play a dominant role in
setting $Y_e$ just before nucleosynthesis starts. This in turn strongly impacts
nucleosynthesis in the ejecta, since the electron fraction is the determining
factor in whether or not a robust $r$-process occurs. In particular, the dynamics
of the MHD explosion influence the conditions under which the electron fraction
of ejected material is set, since the dynamics determine when $\tau_{weak}$ becomes
close to the dynamical timescale and when weak interactions freeze out. In
Fig.~\ref{fig:ye_cmp}, we show the evolution of $Y_e$, $Y_{e, \beta}$, $\tau_{weak}$,
and $\tau_d$ for representative particles from the three simulations. The four
colored lines indicate the $Y_e$ evolution as obtained from the nuclear reaction
network calculation using four constant neutrino luminosities. In these
calculations we assume $L_{\nu_e} = L_{\nu_{\bar{e}}}$ and constant mean neutrino
energies $\langle \epsilon_{\nu_e}\rangle = 10\, \mathrm{MeV}$ and $\langle
\epsilon_{\bar{\nu_e}} \rangle = 14\, \mathrm{MeV}$. The solid black lines show
the $Y_e$ evolution using the neutrino luminosities as obtained from the tracer
particles. There is an initial decrease in $Y_e$ as material is advected inward
and to higher density, causing $Y_{e,eq}$ to go down and $\tau_{weak}$ to decrease. 
Except in a limited number of cases, $\tau_d <
\tau_{weak}$ and $\beta$-equilibrium is never obtained, although $Y_e$ is
always moving towards $Y_{e,\beta}$. Then, as the particle moves outward with
the jet and evolves toward lower density, the
electron capture rate is reduced and neutrino captures begin to dominate the
weak reaction rates, causing $Y_e$ to increase (this occurs before $0.1
\textrm{s}$ in all of the plots). Finally, once the temperature reaches $T
\approx 5 \, \textrm{GK}$, $r$-process nucleosynthesis begins and
$\beta^-$-decays of heavy nuclei cause $Y_e$ to increase. The initial value of
$Y_e$ relevant at the start of $r$-process nucleosynthesis is seen as the
plateau near $1 \textrm{s}$ in Fig.~\ref{fig:ye_cmp}.
 
In the most energetic model, B13, the jet is formed at high densities very soon
after bounce. As the jet propagates out, it entrains collapsing material that
has not fully deleptonized, trapped neutrinos, and reached weak equilibrium.
Therefore, after the ejecta begins to move out to larger radii and smaller
densities, its $Y_e$ still goes down due to electron captures trying to move the
material toward the lower $Y_e$ predicted by neutrino-free $\beta$-equilibrium.
The particles in B13 do not reach their minimum $Y_e$ at their maximum density,
but rather continue to experience electron captures that drive $Y_e$ down toward
its neutrino-free $\beta$-equilibrium value. This generally pushes the ejected
material in B13 to small $Y_e$ values ($Y_e \simeq 0.15$).

In model B12-sym, the evolution differs since it takes $\simeq 20-30\,\mathrm{ms}$
before a jet explosion is launched and the propagation of the jet is slower than
in model B13. Tracer particles that accrete toward the PNS reach
slightly lower maximum densities than in simulation B13, but these particles
reach their lowest $Y_e$ at their maximum
density. As particles get ejected in the outflow, they evolve to higher $Y_e$
due to neutrino interactions. This effect is more pronounced for higher
neutrino luminosities and is enhanced compared to simulation B13 due to the
longer dwell time of particles in the vicinity of the PNS. 

In model B12, the explosion dynamics are drastically different than in models
B13 and B12-sym. The shock starts to expand only after $\simeq 80\, \mathrm{ms}$.
Additionally, particles do not accrete to minimum radii as low as in simulations B13 and
B12-sym, hence they reach smaller maximum densities before being ejected.
The eventually-ejected material comes close to reaching $\beta$-equilibrium
when it reaches its maximum density. Since this maximum density is lower
than the maximum densities encountered in B12-sym, the minimum $Y_e$ reached by
the ejected B12 material is systematically higher. The propagation speed of
the explosion is slower than in models B13 and B12-sym. However, the dwell
time of particles in the vicinity of the PNS before being
ejected is similar to that of model B12-sym. This is due to a similar ejection
speed in the initially forming outflow near the PNS. It is only
the shock surface itself that propagates at slower expansion speeds as the
outflow material spirals away from the rotation axis. As the ejected material
interacts with neutrinos, it evolves to higher $Y_e$ values. This rise in $Y_e$
is similar to the evolution in simulation B12-sym.

\begin{figure}[t]
\includegraphics[width=0.48\textwidth]{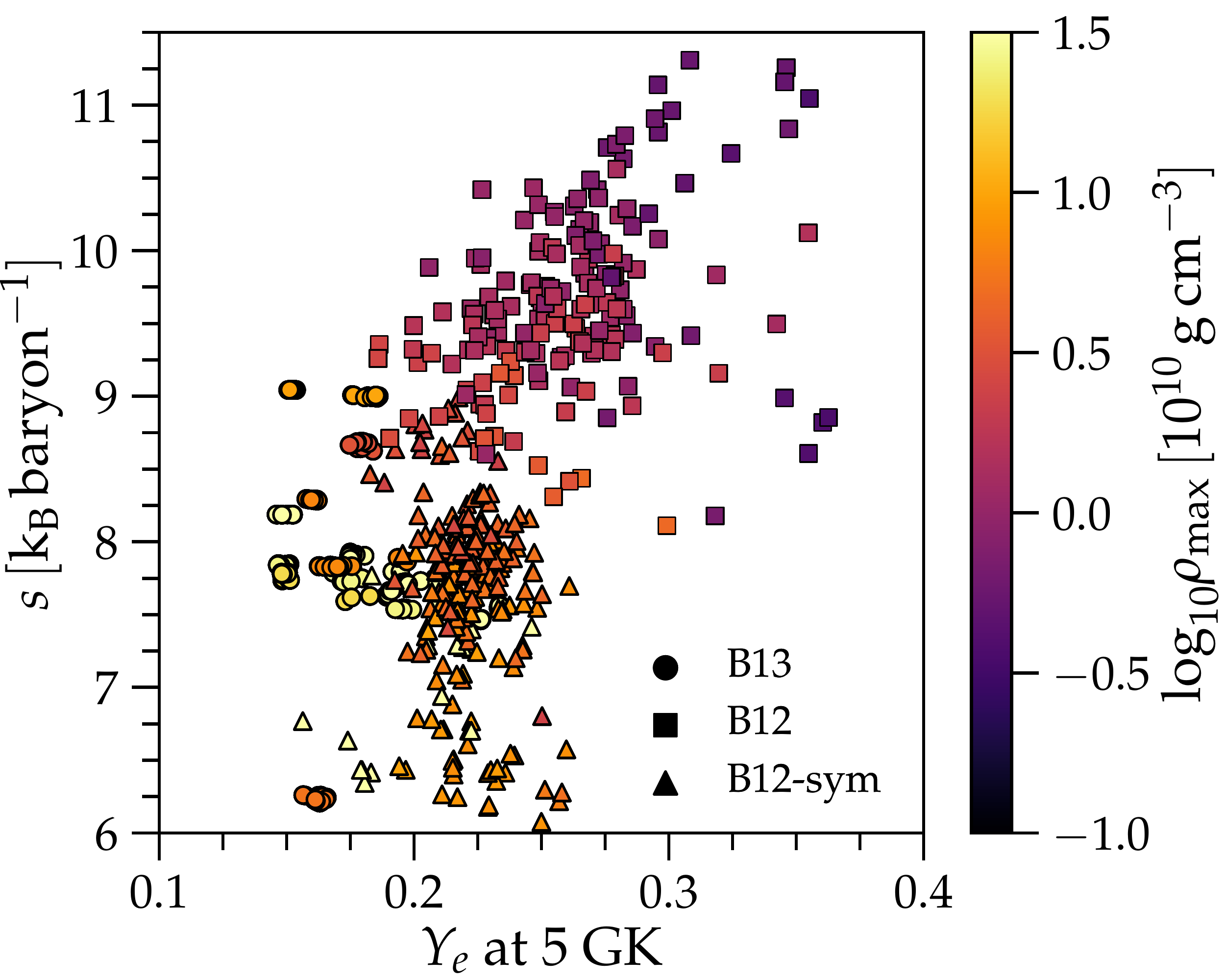}
\caption{Scatter plot of the electron fraction $Y_e$ at $T = 5\, \mathrm{GK}$ (x-axis) and specific entropy $s$ (y-axis) for select
particles from simulations B13 (circles), B12-sym (triangles), and B12 (squares). The symbols
are color-coded with the maximum density $\rho_{\mathrm{max}}$ reached.} 
\label{fig:ye_rho_max_scatter} 
\vspace{0.5cm} 
\end{figure}

We show selected particles from simulations B13, B12-sym, and B12 as a scatter
plot in $Y_e$ at $T=5\, \mathrm{GK}$ and specific entropy in
Fig.~\ref{fig:ye_rho_max_scatter}. This figure illustrates the behavior
described above for individual tracer particles. The symbols for each
particle are color coded with the maximum density reached. For simulation B13,
particles reach the highest densities as they reach the smallest minimum radii.
The $Y_e$ values at the time when the particles last exceed a temperature of 5
GK (approximately the temperature threshold for $r$-process nucleosynthesis)
are peaked at low $Y_e \simeq 0.2$. The entropy values for the particles are
similar to those for simulation B12-sym but lower than for simulation B12. The
low $Y_e$ values for simulation B13 are almost exclusively set by
$\beta$-equilibrium since neutrino irradiation has less of an effect on the
$Y_e$ distribution even for high neutrino luminosities since material gets ejected
very rapidly and efficiently. In model B12-sym, particles are at similar
entropy but lower maximum densities and higher $Y_e$ values at $T=5\,
\mathrm{GK}$ compared to simulation B13. The $Y_e$ values at $T=5\,
\mathrm{GK}$ for the particles from simulation B12-sym are set by two effects.
First, the $\beta$-equilibrium $Y_e$ values for these particles at lower
densities are higher than for the particles at higher densities in simulation
B13. Second, the dwell time for particles in the vicinity of the PNS is an
order of magnitude longer in simulation B12-sym than in B13. This causes the
$Y_e$ values of the particles to shift to higher values as they cool to $T= 5\,
\mathrm{GK}$. In the full 3D simulation B12, particles are at the lowest
maximum densities and highest entropies. Their $Y_e$ values in
$\beta$-equilibrium are therefore higher than for both B13 and B12-sym. The
shift in the $Y_e$ distribution as the particles evolve toward $T\simeq 5\
\mathrm{GK}$ is similar to the evolution in simulation B12-sym.  This is caused
by the similar dwell time of material at small radii before being ejected in
the outflow and hence a similar amount of neutrino interactions. The
distribution of $Y_e$ values for particles from simulation B12 is considerably
wider than for simulations B13 and B12-sym.

In addition to the high density lepton captures, neutrino captures at lower
density can also impact $Y_e$ at the beginning of nucleosynthesis.
We parameterize the neutrino luminosities for the network calculation to
determine how much of an impact uncertainties in our neutrino transport
approximation have on the nuclear network calculation. In all simulations,
higher neutrino luminosities push the particle $Y_e$ values more quickly toward
the higher end. This is particularly pronounced for neutrino luminosities
$L_{\nu} = 10^{52}\, \mathrm{erg\, s^{-1}}$ and $L_{\nu} = 10^{53}\,
\mathrm{erg\, s^{-1}}$. The neutrino luminosities recorded from the tracer
particles peak at a few $L_{\nu} = 10^{52}\, \mathrm{erg\, s^{-1}}$ (see
Fig.~\ref{fig:lum_cmp}) and are bracketed by the $L_{\nu} = 10^{52}\,
\mathrm{erg\, s^{-1}}$ and $L_{\nu} = 10^{53}\, \mathrm{erg\, s^{-1}}$ constant
luminosity cases. 

\subsection{Ejecta composition}
\label{sec:ejecta}

\begin{figure*}[t]
\includegraphics[width=0.32\textwidth]{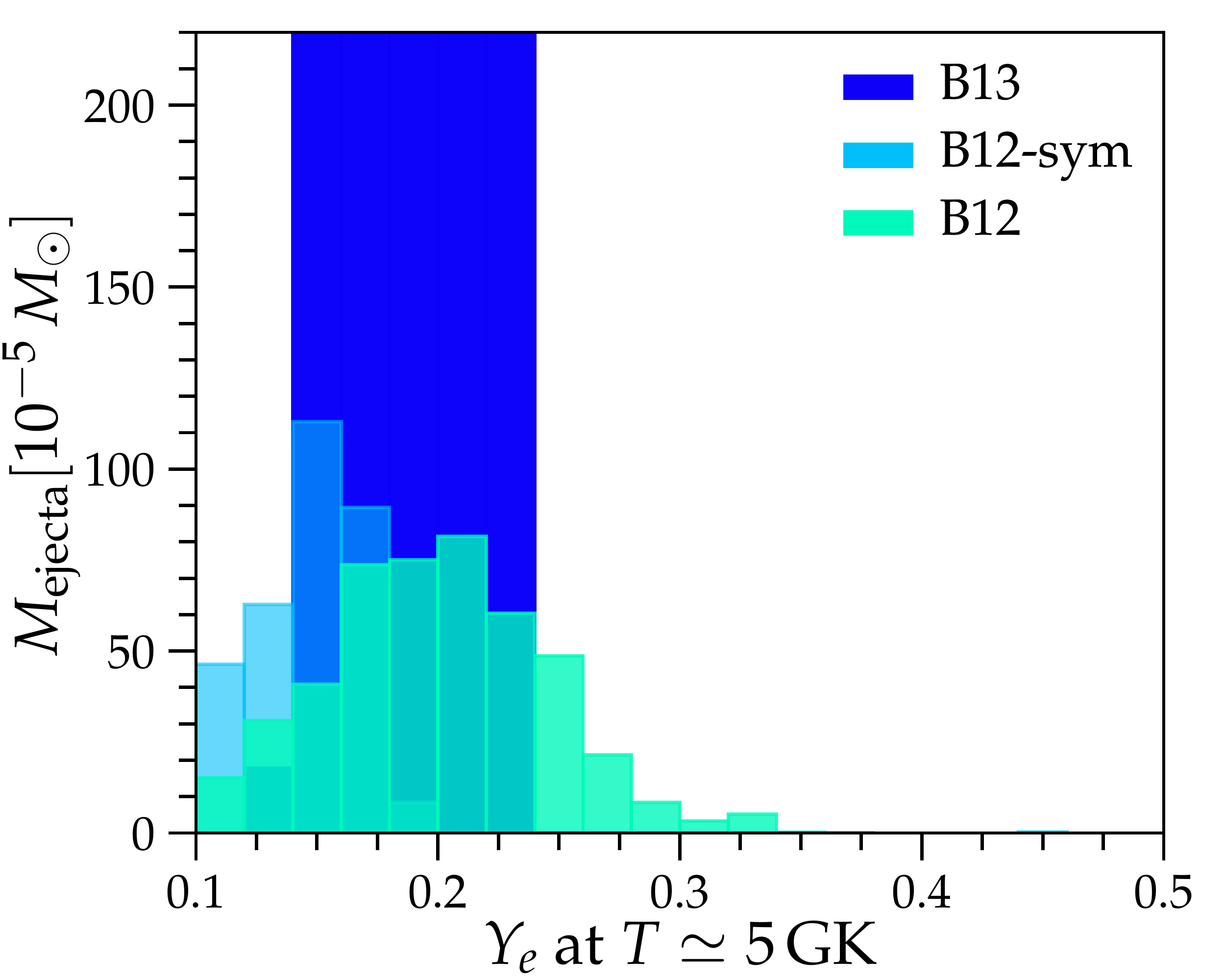}
\includegraphics[width=0.32\textwidth]{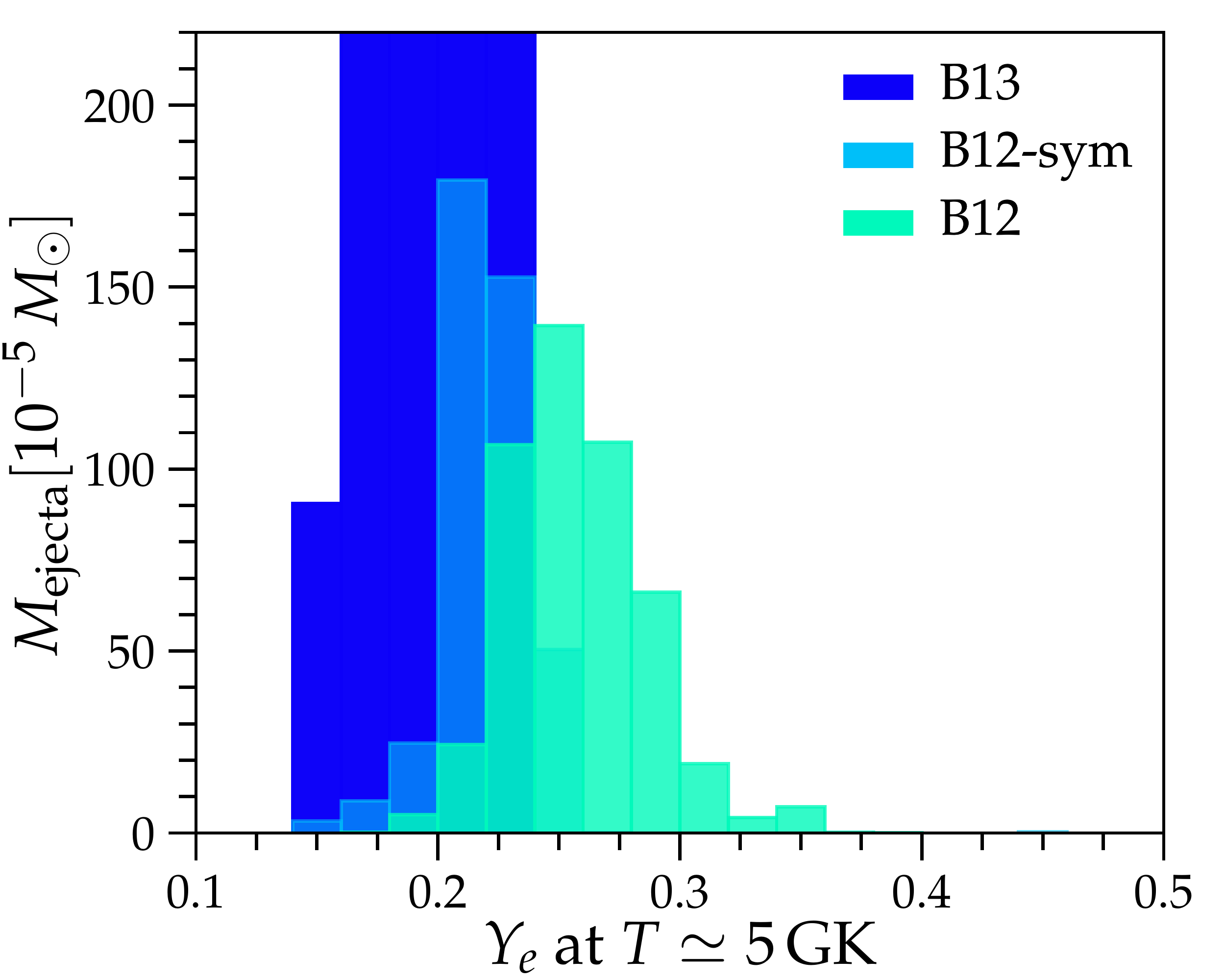}
\includegraphics[width=0.32\textwidth]{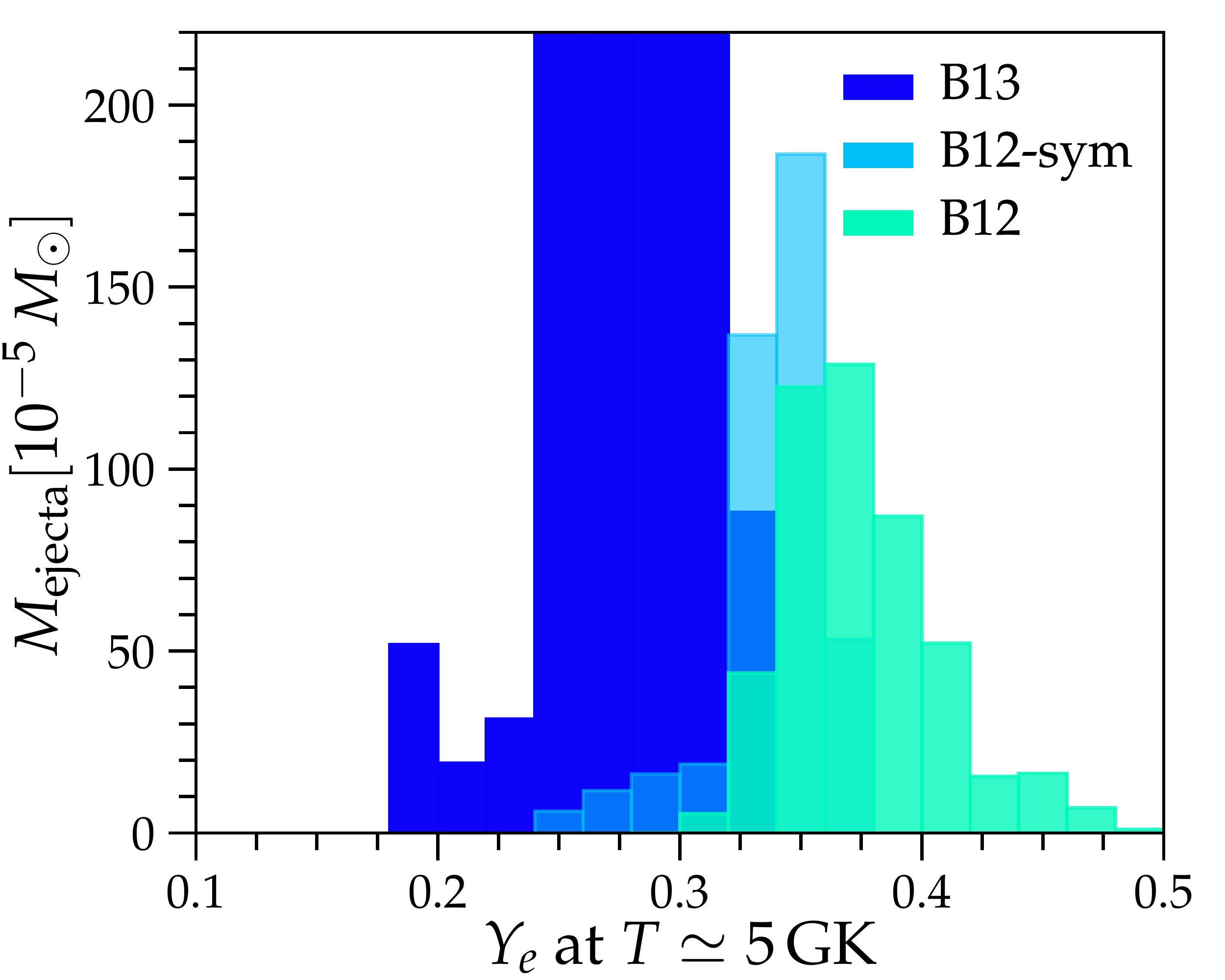}
\caption{$Y_e$ histograms when the particles are above a temperature of T=5 GK
	for the last time. We show simulation B13 (dark blue), B12-sym (cyan),
	and B12 (green). The left panel shows results obtained without taking
	neutrino luminosities into account for the network calculation. The
	center panel shows results obtained with constant neutrino luminosities
	$L_{\nu_e} = L_{\bar{\nu_e}} = 10^{52}\, \mathrm{erg\, s^{-1}}$,
	and the right panel shows results obtained using the luminosities
recorded from the tracer particles. We bin $Y_e$ in intervals of 0.02 and weigh
the $Y_e$ statistics with the mass of the ejected particles.} 
	\label{fig:ye_hist} 
\vspace{0.5cm} 
\end{figure*}
\begin{figure*}[t]
\includegraphics[width=0.32\textwidth]{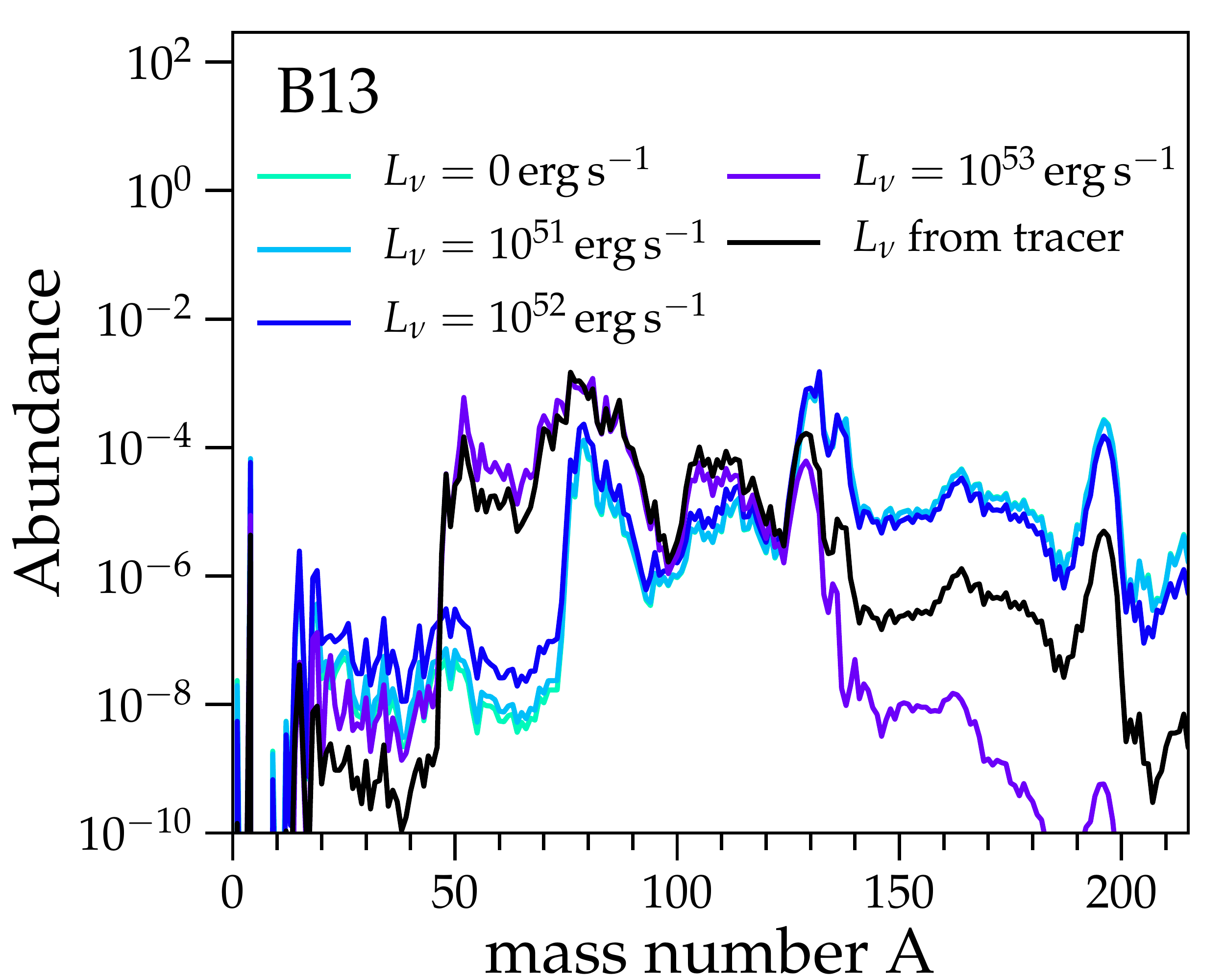}
\includegraphics[width=0.32\textwidth]{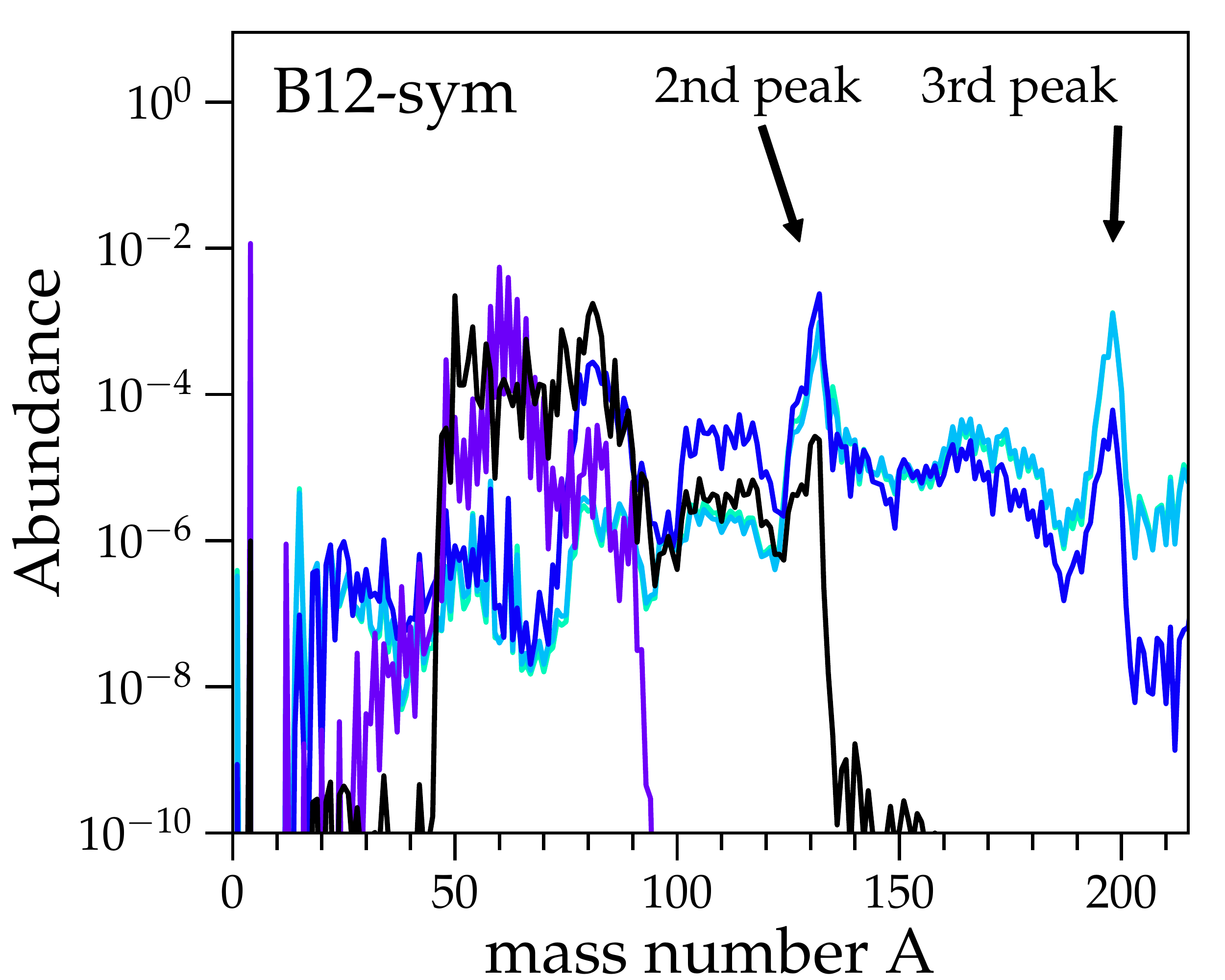}
\includegraphics[width=0.32\textwidth]{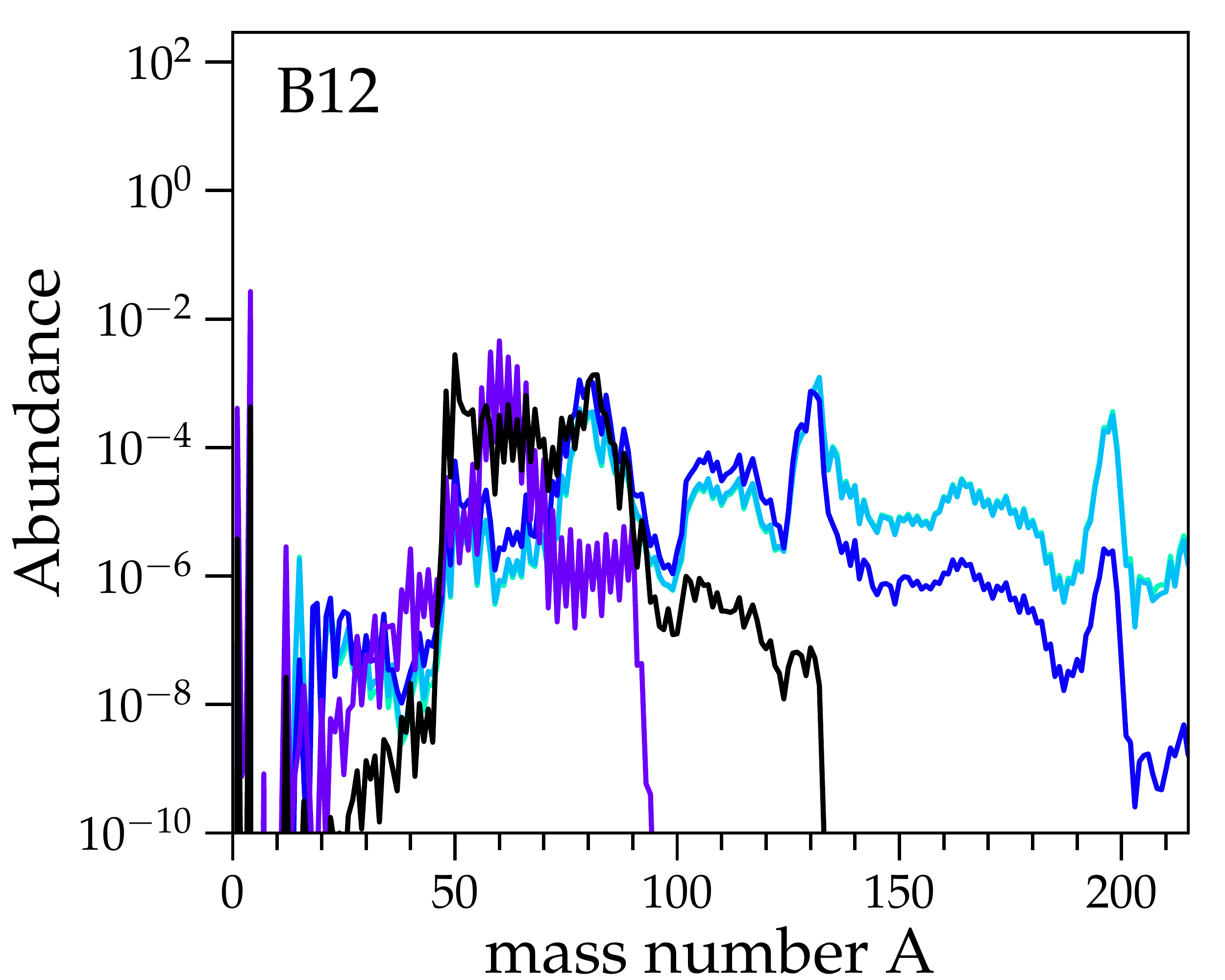}
\caption{Fractional abundance as a function of mass number $A$ for
	models B13 (left), B12-sym (center), and B12 (right). Differently
	colored lines indicate results obtained with different constant neutrino
	luminosities in the nuclear reaction network calculation. Black lines
	show the results obtained when using the neutrino luminosities as
	recorded from the tracer particles in the simulations. For model B13,
	neutrino luminosities up to $L_{\nu_e} = L_{\bar{\nu_e}} = 10^{52}
	\mathrm{erg\, s^{-1}}$ produce a robust second and third peak $r$-process
	pattern. Starting from a neutrino luminosity of $L_{\nu_e} = L_{\bar{\nu_e}}
	= 10^{53} \mathrm{erg\, s^{-1}}$ and the neutrino luminosity from the tracer
	particles material beyond the second peak is reduced in abundance.
	This trend is continued in models B12-sym and B12, but with a
	reduction in abundance of nuclei beyond the second peak
	starting at lower and lower neutrino luminosities. For model B12, only
	$L_{\nu_e} = L_{\bar{\nu_e}} = 10^{51} \mathrm{erg\, s^{-1}}$ still
	produces a robust $r$-process abundance pattern.}
\label{fig:abund_all} 
\vspace{0.5cm} 
\end{figure*}

The ejecta properties vary significantly between the simulations. Most
important for the $r$-process nucleosynthetic signature of the explosion is how
neutron-rich the ejected material is. In Fig.~\ref{fig:ye_hist}, we show the
distribution of the electron fraction $Y_e$ for all particles in the ejected
material when the temperature for the particles is last above 5 GK. This is
representative of $Y_e$ at the beginning of neutron-capture nucleosynthesis and
leads to different ejecta properties between jet explosions (simulations B13
and B12-sym) and the 3D dual-lobe explosion (B12). We show results for both the
leakage neutrino luminosities and our assumed constant neutrino luminosities.
In the case of the leakage neutrino luminosities, the luminosities are also
assumed constant in the network calculation after the end of the tracer
particle data. 

For zero neutrino luminosities, the distributions for all simulations are peaked
at $Y_e \lesssim 0.2$. B12-sym is peaked at lower $Y_e \simeq 0.15$ than B12 at
$Y_e \simeq 0.21$. The distribution for B12 is significantly broader than for
B12-sym and B13. There are more particles at low $Y_e$ values for simulation
B12-sym than for B12. This is caused by particles reaching higher densities
before they get turned around and swept up in the outflow (see
Fig.~\ref{fig:ye_rho_max_scatter}). In the $10^{52}\, \mathrm{erg\, s^{-1}}$
luminosity case, neutrino interactions shift the distributions to higher $Y_e$
for all simulations. For model B13, where the dwell time of particles in the
neutrino field is a factor $\simeq$ 10 shorter than in models B12-sym and B12, this
shift is not large, but for simulations B12-sym and B12 the distributions are
shifted by almost $\simeq 0.1$. As a result, there is effectively no material
at $Y_e \lesssim 0.2$ for simulation B12-sym, and no material below $Y_e
\simeq0.22$ for simulation B12. The results obtained with the neutrino
luminosities from the tracer particles show this effect even more clearly.
Here, the effect of neutrino interactions is large enough that even the $Y_e$
distribution for simulation B13 is shifted to values of $Y_e \gtrsim 0.2$,
the distribution for B12-sym is now centered at $Y_e \simeq 0.34$, and the
$Y_e$ distribution for simulation B12 is shifted to $Y_e \simeq 0.36$. 

The variations in the distribution of $Y_e$ have consequences for the eventual
nucleosynthesis, since one must have $Y_e \lesssim 0.25$ to make the 
third $r$-process peak \citep{lippuner:15}. Figure~\ref{fig:abund_all} shows
abundance patterns for all three simulations, B13, B12-sym, and B12. We show
the fractional abundance pattern averaged over all particles in the ejecta as a
function of mass number $A$. 

If no neutrino luminosities are taken into account in the nucleosynthesis
calculation, we find a robust $r$-process pattern in all three simulations. This
is also true for constant neutrino luminosity of $L_{\nu_e} = L_{\bar{\nu_e}}
= 10^{51} \mathrm{erg\, s^{-1}}$. For neutrino luminosity $L_{\nu_e} =
L_{\bar{\nu_e}} = 10^{52} \mathrm{erg\, s^{-1}}$ all simulations still show a
robust second $r$-process peak. B13 still has robust third peak
abundances, while B12-sym and B12 have reduced abundances in their third peaks
(with B12 seeing the larger reduction). For neutrino luminosity of $L_{\nu_e} =
L_{\bar{\nu_e}} = 10^{53} \mathrm{erg\, s^{-1}}$ none of the simulations show
significant amounts of material synthesized beyond $A=135$. In all
simulations, the reduction in fractional abundance beyond $A=135$ is
accompanied by an overproduction of nuclei with $A<135$ compared to the lower
neutrino luminosity cases.

\begin{figure}[t]
\includegraphics[width=0.48\textwidth]{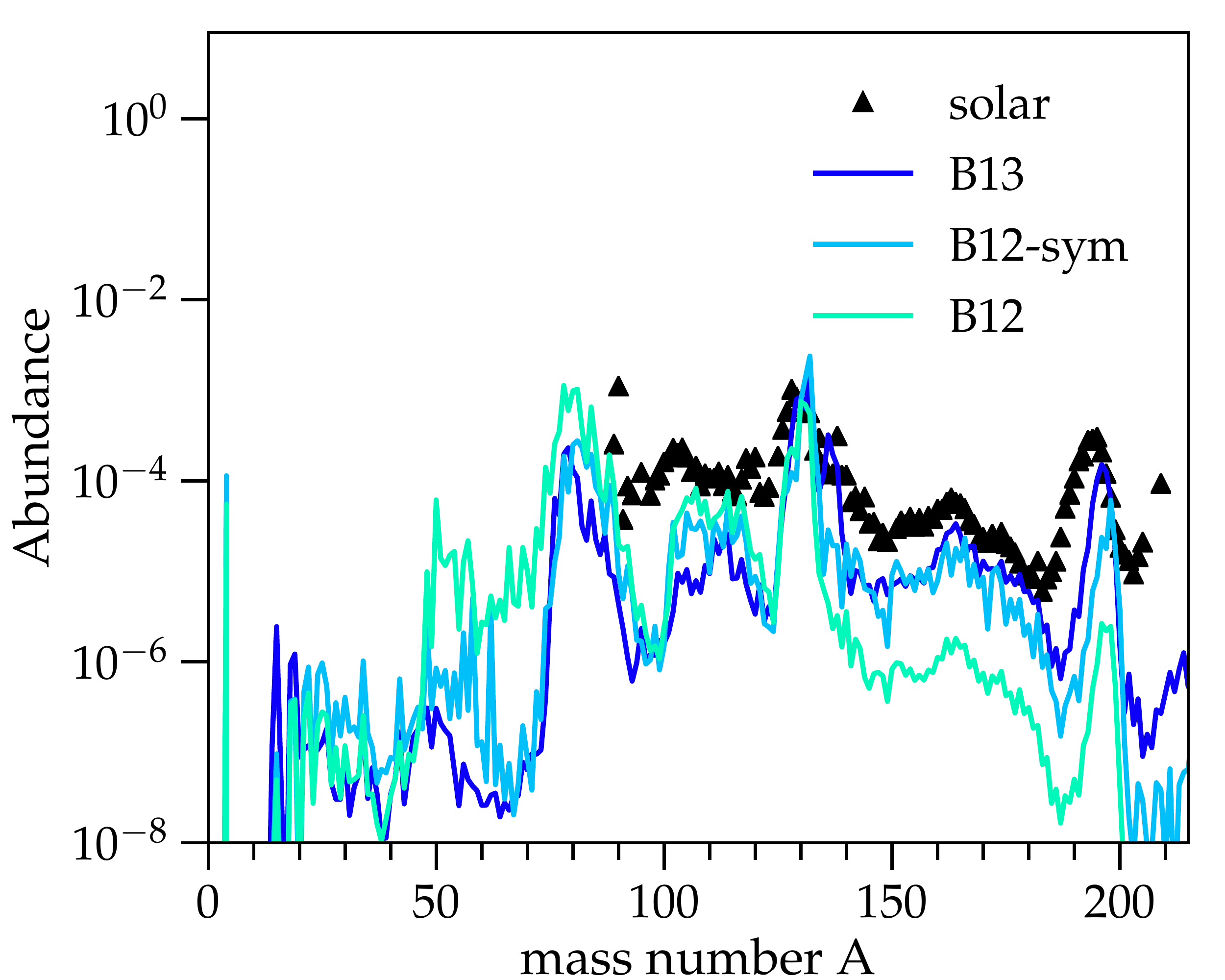}
\caption{Fractional abundance pattern as a function of mass number $A$ for
	models B13, B12-sym and B12. Blue, light blue, and light green show models B13,
	B12-sym, and B12, respectively, for a constant neutrino luminosity
	of $L_{\nu_e} = L_{\bar{\nu_e}} = 10^{52} \mathrm{erg\, s^{-1}}$ for both
	electron and electron antineutrinos in the nuclear reaction network
	calculation. Black markers indicate the solar abundance pattern scaled
	to match the second $r$-process peak ($A=135$) for simulation B13.
	Model B13 reproduces the solar abundance pattern reasonably well, while
	model B12-sym underproduces third $r$-process peak ($A=195$) material
	by more than an order of magnitude. In model B12, all nuclei beyond the
	second $r$-process peak are reduced in abundance by a factor of $\sim 100$.}
\label{fig:abund_cmp} 
\vspace{0.5cm} 
\end{figure}

The abundance patterns calculated with the neutrino luminosities as recorded
from the tracer particles fall in between the $L_{\nu_e} = L_{\bar{\nu_e}} =
10^{52} \mathrm{erg\, s^{-1}}$ and $L_{\nu_e} = L_{\bar{\nu_e}} = 10^{53}
\mathrm{erg\, s^{-1}}$ constant luminosity cases. For simulation B13, material
beyond $A>135$ is reduced by a factor 10 relative to the $L_\nu=0$ case, but
for simulations B12-sym and B12, the results with luminosities as recorded from
the tracer particles follow the $L_{\nu_e} = L_{\bar{\nu_e}} = 10^{53}
\mathrm{erg\, s^{-1}}$ case closely. There is no or very little second or
third peak $r$-process material synthesized.

For a more direct comparison we show nucleosynthesis calculations with constant
neutrino luminosities of $L_{\nu_e} = L_{\bar{\nu_e}} = 10^{52} \mathrm{erg\,
s^{-1}}$ for simulations B13, B12-sym, and B12 in Fig.~\ref{fig:abund_cmp}. 
While model B13 matches the solar abundance pattern well, model B12-sym falls
short in the amount of material synthesized beyond $A\simeq170$ by a factor of
a few. For third-peak $r$-process material, the reduction in abundance between
models B13 and B12-sym is slightly more than a factor of 10. For model B12,
the reduction in material beyond the second peak is even more severe. Material
beyond $A=135$ is underproduced by two orders of magnitude with respect to
simulation B13 and the solar abundance pattern. This underproduction is
accompanied by an overproduction of nuclei with mass numbers $50 \leq A \leq
80$. 

\begin{table}
\centering
\begin{threeparttable}
\caption{Total and $r$-process ejecta masses (material with $ 120 \leq A \leq
	249$) for the three simulations B13, B12-sym, and B12 for the four constant
neutrino luminosities and the neutrino luminosities as obtained from the tracer
particles. Masses are in solar masses $M_{\odot}$.}
\begin{tabular}{lccc}
    \hline
    \hline
    Simulation & B13 & B12-sym & B12\\ 
    \hline
    $M_{\mathrm{ej,tot}}\, \mathrm{[M_{\odot}]}$  & 0.0356 & 0.0043 & 0.0048\\ 
    $M_{\mathrm{ej,r}}$ \quad $L_{\nu} = 0\, \mathrm{erg\, s^{-1}}\, \mathrm{[M_{\odot}]}$ & 0.0337 & 0.0042 & 0.0038  \\ 
    $M_{\mathrm{ej,r}}$ \quad $L_{\nu} = 10^{51}\, \mathrm{erg\, s^{-1}}\, \mathrm{[M_{\odot}]}$& 0.0336 & 0.0042 & 0.0037  \\ 
    $M_{\mathrm{ej,r}}$ \quad $L_{\nu} = 10^{52}\, \mathrm{erg\, s^{-1}}\, \mathrm{[M_{\odot}]}$& 0.0320 & 0.0034 & 0.0018   \\ 
    $M_{\mathrm{ej,r}}$ \quad $L_{\nu}$ from tracer $\, \mathrm{[M_{\odot}]}$ & 0.0038 & $5.4\times 10^{-5}$ & $4.0\times 10^{-7}$ \\ 
    $M_{\mathrm{ej,r}}$ \quad $L_{\nu} = 10^{53}\, \mathrm{erg\, s^{-1}}\, \mathrm{[M_{\odot}]}$& 0.0012 & 0.0 & 0.0  \\ 
    \hline                                                              
\end{tabular}
\label{tab:ejectedmass}
\vspace{0.2cm}
\end{threeparttable}
\end{table}

Table~\ref{tab:ejectedmass} summarizes the mass of the total and $r$-process
ejecta material for models B13, B12-sym, and B12. The ejecta mass for
simulation B13 is an order of magnitude larger than for simulations B12-sym and
B12. This is due to the immediate jet launch after core bounce and the
propagation speed of $v \simeq 0.15 c$. All of the ejected mass measurements
are only lower bounds on the total ejecta mass, since it is still increasing at
the end of each of the simulations. Our most optimistic neutrino luminosity
scenario that is still within the uncertainty of the Leakage luminositites from
the tracer partcles is constant neutrino luminosity of $10^{52}\, \mathrm{erg
s^{-1}}$. For this luminosity, the $r$-process ejecta mass in model B13 is
comparable to that found in \cite{winteler:12} and \cite{nishimura:15}.
For neutrino luminosities taken from the tracer particles and for constant
neutrino luminosities of $10^{53}\, \mathrm{erg \, s^{-1}}$, the $r$-process
ejecta mass is reduced by an order of magnitude. In simulations B12-sym and
B12, the $r$-process ejecta mass for our most optimistic scenario is already an
order of magnitude smaller than for the same neutrino luminosity in simulation
B13. For neutrino luminosities taken from the tracer particles and for constant
neutrino luminosities of $10^{53}\, \mathrm{erg \, s^{-1}}$, the $r$-process
ejecta mass is effectively zero. 

\section{Discussion}
\label{sec:discussion}

We have studied $r$-process nucleosynthesis from a set of 3D CCSNe simulations.
Our models include a full 3D simulation with a precollapse magnetic field of
$10^{13}\,\mathrm{G}$ (B13) that is similar in dynamics to the simulation
presented in \cite{winteler:12}, a 3D simulation set up to be identical in
dynamics to an axisymmetric simulation with a precollapse magnetic field of
$10^{12}\, \mathrm{G}$ (B12-sym) that is similar to the prompt axisymmetric jet
explosions in \cite{nishimura:15}, and a full 3D simulation with a precollapse
magnetic field of $10^{12}\, \mathrm{G}$ (B12) as in \cite{moesta:14b}. In our
nuclear reaction network calculations we have included weak interactions to
account for interaction of material with neutrinos emitted from the PNS. We
have specifically used both parameterized constant neutrino luminosities and
the recorded neutrino luminosities from the tracer particles in the
simulations. 

Our results show that the nucleosynthetic signature of 3D magnetorotational
CCSNe depends on the detailed dynamics of the jet and the neutrino emission
from the PNS. Our 3D simulations that include a factor 10 lower initial
magnetic field differ fundamentally from what was anticipated based on either
axisymmetric simulations~\citep{nishimura:15} or 3D simulations of very highly
($B \geq 5\times 10^{12}\, \mathrm{G}$) magnetized progenitor
cores~\citep{winteler:12}.

We find that weak interactions in the nuclear reaction network calculations
change the nucleosynthetic signatures of all simulations. Including no
neutrino luminosities in the network calculation based on simulation B13
produces a robust $r$-process abundances consistent with the observed solar
abundance pattern and with what \cite{winteler:12} found.  Starting with
neutrino luminosities of $5\times 10^{52}\, \mathrm{erg\, s^{-1}}$, $r$-process
material beyond the second peak is reduced in abundance by a
factor of a few, and by an order of magnitude for larger neutrino luminosities.
For simulation B12-sym, the reduction in synthesized nuclei beyond the second
$r$-process peak starts at neutrino luminosities of $10^{52}\, \mathrm{erg\,
s^{-1}}$ but matters mostly for third peak $r$-process nuclei. For simulation
B12, the reduction in abundance of nuclei beyond the second peak is
consistently at least a factor ten compared to the lower neutrino luminosity
calculations. The neutrino luminosities recorded by the tracer particles
typically are a few $10^{52}\, \mathrm{erg\, s^{-1}}$ after the initial
neutronization burst has subsided after $\sim 20\, \mathrm{ms}$, and hence fall
in between the constant luminosity cases of $10^{52}\, \mathrm{erg\, s^{-1}}$
and $10^{53}\, \mathrm{erg\, s^{-1}}$. Acknowledging a factor of 
$\sim$ 2 uncertainty in the neutrino luminosities in our simulations, we
compare a neutrino luminosity of constant $10^{52}\,\mathrm{erg\, s^{-1}}$
between the three simulations B13, B12-sym, and B12 and the solar abundance
pattern as our most optimistic neutrino luminosity scenario. We find a robust
second and third peak abundance pattern only for simulation B13.  Simulation
B12-sym shows an underproduction of nuclei beyond $A=170$ by a factor of a few.
For the full 3D simulation B12, we find that nuclei beyond the second
$r$-process peak are underproduced by a factor of 100 compared to solar
abundances. 

Our results show that the realistic 3D dynamics of magnetorotationally-driven CCSNe
change their $r$-process nucleosynthetic signatures. The different explosion
dynamics lead to ejecta material probing different regions of the engine
driving the explosion. In simulation B13, material from the smallest radii gets
entrained in the outflow, while material in the outflows of simulations B12-sym
and B12 originates at larger radii and lower densities. This leads to less
neutron-rich material being entrained in the outflows for simulation B12. In
addition, the dwell time of ejecta material in the vicinity of the PNS for
simulations B12-sym and B12 is a factor of $\simeq$ 10 longer than for
simulation B13. This causes the $Y_e$ distribution of the ejecta at the onset
of $r$-process nucleosynthesis to shift to higher $Y_e$. This is especially
true for simulation B12, for which the distribution of $Y_e$ in the ejecta at
the onset of $r$-process nucleosynthesis is peaked at $\sim 0.28$ and is
broader than for simulations B13 and B12-sym. For the full 3D dynamics of the
explosion in simulation B12, the mass of ejected $r$-process material is an
order of magnitude smaller even for our most optimistic scenario of constant
neutrino luminosity of $10^{52}\, \mathrm{erg s^{-1}}$. For neutrino
luminosities obtained from the tracers particles, the ejected $r$-process mass
is only $\sim 10^{-7} M_{\odot}$. Most importantly, third peak material is a
factor 100 less abundant when compared with perhaps unrealistic jet explosions
like simulations B13 and B12-sym. 

Our results suggest that the only viable channel for a robust $r$-process
pattern is via an immediate jet explosion at core bounce. Such an explosion is
extremely effective at funneling material into the jet-driven outflow leaving
little time for weak interactions to push the ejected material to higher
electron fraction values. For this case and as in \cite{winteler:12}, we find
a robust $r$-process abundance pattern consistent with observed solar
abundances. Immediate jet launch at core bounce requires $10^{16}\,\mathrm{G}$
of large-scale toroidal field and poloidal field of similar strength to
stabilize the outflow against the kink instability. This field can be generated
via amplification by the MRI and a dynamo process, but the amplification will
take at least 10 spin periods of the PNS~\citep{moesta:15}. For realistic
precollapse iron cores with magnetic fields not in excess of
$\sim$$10^{8}-10^{9}\,\mathrm{G}$, this amplification timescale is even longer.
In addition, the MRI and dynamo action will likely saturate at field strengths
of no more than $\sim$$10^{15}\, \mathrm{G}$~\citep{rembiasz:16} for both the
poloidal and toroidal components, falling short of the required ultra-strong
poloidal field required to stabilize the jet. Therefore, the magnetic field in
simulation B13 (and similarly the fields in the simulation presented in
\citealt{winteler:12}) cannot be assumed to be delivered by this amplification
channel. The only viable channel to achieve these field strengths at core
bounce is thus the collapse of iron cores with sufficiently strong
($B\gtrsim$$10^{13}\,\mathrm{G}$) precollapse poloidal fields, which are 
likely unrealistic.

The reduced abundance of ejecta material beyond the second peak from our
simulations changes the predicted yield of $r$-process material per event for
magnetorotational supernovae. This will have to be taken into account when
studying galactic chemical evolution and the role of magnetorotational
supernovae in early $r$-process enrichment. It is particularly important
given recent evidence that neutron star mergers may not be able to explain
$r$-process enrichment in the lowest metallicity stars \citep[e.g.][]{casey:17}
and evidence for multiple distinct enrichment channels in dwarf
galaxies~\citep{tsujimoto:17}.

\section*{Acknowledgments}

The authors would like to thank D.~Kasen, E.~Quataert, and D.~Radice for
discussions. This research was partially supported by NSF grants AST-1212170,
CAREER PHY-1151197, OAC-1550514 and OCI-0905046. PM acknowledges support by
NASA through Einstein Fellowship grant PF5-160140. The simulations were carried
out on XSEDE resources under allocation TG-AST160049 and on NSF/NCSA BlueWaters
under NSF award PRAC OCI-0941653. This paper has been assigned Yukawa Institute
for Theoretical Physics report number YITP-17-129 and LANL Report number
LA-UR-17-31278.

\bibliography{bibliography/jet_references,bibliography/bh_formation_references,bibliography/gw_references,bibliography/sn_theory_references,bibliography/grb_references,bibliography/nu_obs_references,bibliography/methods_references,bibliography/eos_references,bibliography/NSNS_NSBH_references,bibliography/stellarevolution_references,bibliography/nucleosynthesis_references,bibliography/gr_references,bibliography/nu_interactions_references,bibliography/sn_observation_references,bibliography/populations_references,bibliography/pns_cooling_references,bibliography/spectral_photometric_modeling_references,bibliography/cs_hpc_references,bibliography/numrel_references,bibliography/radiation_transport_references,bibliography/mhd_references}

\end{document}